\documentclass[twocolumn]{article}

\usepackage[title]{appendix}

\usepackage{subfigure}
\usepackage[pdftex]{graphicx}
\usepackage{epstopdf}
\usepackage{multirow,setspace}

\usepackage[table]{xcolor}
\usepackage{amsmath}
\usepackage{amsthm}

\newtheorem{theorem}{Theorem}
\DeclareMathOperator{\arccot}{arccot}
\usepackage{cite}

\begin{document}

\title{Discrete Linear Canonical Transform Based on Hyperdifferential
Operators} 

\author{Aykut~Ko\c{c},
        Burak~Bartan,
        and~Haldun~M.~Ozaktas,
\thanks{Aykut Ko\c{c} (Corresponding Author) is with ASELSAN Research Center, Ankara, Turkey, e-mail: aykutkoc@aselsan.com.tr.}
\thanks{Burak Bartan is with Electrical Engineering Department, Stanford University, Stanford, California}
\thanks{Haldun M. Ozaktas is with Electrical Engineering Department, Bilkent University, TR-06800 Bilkent, Ankara, Turkey}}

\maketitle

\begin{abstract}
Linear canonical transforms (LCTs) are of importance in many areas
of science and engineering with many applications. Therefore 
a satisfactory discrete implementation is of considerable interest.
Although there are methods that link the samples of the input
signal to the samples of the linear canonical transformed output
signal, no widely-accepted definition of the discrete LCT has been
established. We introduce a new approach to defining the discrete
linear canonical transform (DLCT) by employing operator theory. Operators are abstract entities that can have both continuous and discrete concrete manifestations. Generating the continuous and discrete manifestations of LCTs from the same abstract operator framework allows us to define the continuous and discrete transforms in a structurally analogous manner. By utilizing hyperdifferential operators, we obtain a DLCT matrix which is totally compatible with the theory of the discrete Fourier transform (DFT) and its dual and circulant structure, which makes further analytical manipulations and progress possible. The proposed DLCT is to the continuous LCT, what the DFT is to the continuous Fourier transform (FT). The DLCT of the signal is obtained simply by multiplying the vector holding the samples of the input signal by the DLCT matrix.
\end{abstract}

%

\section{Introduction}
\label{intro}

Linear canonical transforms (LCTs) are a family of linear integral transforms with three parameters,
\cite{wolf79book,ozaktas01book,ozaktas16lctbook,pei02eigenLCT}. The family of LCTs
is a generalization of many important transforms such as the
fractional Fourier transform (FRT), chirp multiplication (CM), chirp
convolution (CC), and scaling operations. For certain values of the
three parameters, the LCT reduces to these transforms or their
combinations. LCTs have several applications in signal processing
\cite{ozaktas16lctbook} and computational and applied mathematics
\cite{davies78integtransappl,koc16lctchapter}, including fast and 
efficient optimal filtering \cite{barshan97optfilt}, radar signal
processing \cite{chen14radar,chen15radar}, speech processing
\cite{qiu13speech}, image representation \cite{koc17eswalct}, and
image encryption and watermarking \cite{singh10encryp,li14watermark,qi15watermark},
to mention a small sample of published works. LCTs have also been
extensively studied for their applications in
optics~\cite{ozaktas01book,siegman86book,bastiaans78wigner,bastiaans79wigner,alieva07lctproperties,bastiaans07abcdclassification,simon00setparaxial},
electromagnetics, and classical and quantum
mechanics~\cite{ozaktas16lctbook,wolf79book,moshinsky73quantLCT,jung82quantcLCT}.

In optical contexts, LCTs are commonly referred to as quadratic-phase
integrals or quadratic-phase systems \cite{bastiaans79wigner,ozaktas06ol}. The
so-called $ABCD$ systems widely used in optics~\cite{hecht01book} are
also represented by linear canonical transforms. They have also been
referred to by other names: generalized Huygens integrals
\cite{siegman86book}, generalized Fresnel transforms
\cite{james96genFresnel,palma97mainABCD}, special affine Fourier
transforms \cite{abe94mainlct,abe94abcd}, extended fractional
Fourier transforms \cite{hua97lct}, and Moshinsky-Quesne
transforms \cite{wolf79book}. 

Two-dimensional (2D) LCTs and complex-parametered LCTs (CLCTs) have
also been discussed in the literature, \cite{rodrigo062DNSLCT,koc10nonsep2D,koc10complexLCT,feng162DLCT}. Bilateral Laplace transforms,
Bargmann transforms, Gauss-Weierstrass transforms,
\cite{wolf79book,wolf74clct1,wolf77clct}, fractional Laplace
transforms, \cite{torre02fracLaplace,sharma09fracLaplace}, and
complex-ordered FRTs
\cite{shih95cFRT,bernardo96cFRT,wang02cFRT,bernardo97cFRTtalbot}
are all special cases of CLCTs.

The establishment of a discrete framework is essential
to the deployment of LCTs in applications. There is considerable work on
discrete or finite forms of fractional Fourier transforms, and, to a
lesser degree, discrete or finite linear canonical transforms. Being
one of the most important special cases of LCTs, discretization and
discrete versions of fractional Fourier transforms have been well studied and
established~\cite{candan00dfrt,wolf05frt,wolf07dfrt,pei97discretefrt,atakishiyev99dfrt,pei99dfrt,pei99dfrt2,zayed99frtsampling,yetik00frtdec,vargasrubio05centered,yeh05dfrt,erseghe99dfrt,barker00dfrt}.

As for the discretization or digital computation of LCTs, there are many approaches present in the literature, \cite{pei00DLCTdef,zhao08rateconversionLCT,stern06whyis,zhang13DLCT,koc08ieee,oktem09DLCT,pei16fast_dlct,ozaktas06ol,campos11fastLCT, hennelly05fastLCT,hennelly05fastLCT2,healy09lctsampling,healy10LCTalgoreview,healy10reevaluation,pei11dlct,zhao13dlct,wei16randomDLCT}. 
Some of these \cite{healy10LCTalgoreview,healy10reevaluation,zhao08rateconversionLCT,stern06whyis,zhang13DLCT,koc08ieee,pei16fast_dlct,ozaktas06ol,campos11fastLCT}
 numerically compute the continuous integral and establish a direct
mapping between the samples of the continuous input function and the
samples of the LCT-transformed continuous output function. The methods in \cite{pei00DLCTdef,zhao08rateconversionLCT,stern06whyis,zhang13DLCT,healy10LCTalgoreview} directly convert the LCT integral to a summation and \cite{koc08ieee,oktem09DLCT,pei16fast_dlct,ozaktas06ol,campos11fastLCT,healy10reevaluation}
make use of decompositions into elementary building blocks. Moreover, some approaches focus on defining a discrete LCT (DLCT),
which can then be used to numerically approximate continuous LCTs,
in the same way that the discrete Fourier transform (DFT) is used to
approximate continuous Fourier transforms \cite{hennelly05fastLCT,hennelly05fastLCT2,healy09lctsampling,pei11dlct,zhao13dlct,wei16randomDLCT,pei00DLCTdef,stern06whyis,oktem09DLCT}.
Algorithms in \cite{pei00DLCTdef,stern06whyis,oktem09DLCT} also numerically
approximate the continuous LCTs in the same way the DFT approximates the
continuous FT. Based on the DLCT definition proposed in \cite{pei00DLCTdef}, Refs.~\cite{healy10LCTalgoreview} and \cite{healy10reevaluation} propose efficient numerical computation algorithms. Ref.~\cite{pei00DLCTdef} also includes a comparison of the properties satisfied by definitions of DLCTs proposed up to that date.

Despite these works, no single definition has been widely
established as the definition of the DLCT. In this paper, we present a different approach based on
hyperdifferential operator theory \cite{ozaktas01book,wolf79book,ozaktas10PoularikasBook,yosida84operationalcalc,nazarathy80operatoralgebra}, to obtain a 
definition of the DLCT. Why do we propose to use operator theory? Most approaches to
discretization are naturally based on sampling of the continuous
entities. However, sampling often does not lead to a clean, discrete transform definition that satisfies operational formulas and exhibits desirable analytical properties such as unitarity and preservation of the group structure. So if our
purpose is not to merely numerically compute a continuous transform,
but to obtain a self-consistent discrete transform definition, it
often turns out to be insufficient. A purely numerical method can compute the continuous transform accurately, but it does not provide us with a definition on which further manipulation can be done, and theoretical progress can build upon. We want a discrete definition that is as analogous to the continuous definition as possible. (This is satisfied by the discrete Fourier transform (DFT) and that is why the DFT is so established.)

How does operator theory help? Operators are abstract entities that can have both continuous and discrete concrete manifestations. Thus if we begin from a continuous entity and can appropriately deduce the abstract operator underlying
that entity, then, that can form a basis for defining its discrete
version. Since both the continuous and discrete versions are based on
the same abstract operator, they can be expected to exhibit similar
structural characteristics and operational properties to the extent possible. The structure
of relationships between different entities can also be preserved and
can be expected to mirror the relationships between the abstract
operators. Thus we can obtain discrete entities that are not merely
numerical approximations, but which exhibit desirable analytical and
operational properties. This is the rationale of the present paper.

Our definition of the discrete LCT will be presented in the form of a
matrix of size $N\times N$ which, upon multiplication, produces the
DLCT of a discrete and finite signal of length $N$, expressed as a
column vector. The main difference from earlier approaches is that the definition is based on hyperdifferential forms of the discrete coordinate multiplication and differentiation operators, which we carefully define so that they are strictly Fourier duals related through the DFT matrix. Our definition provides a self-consistent, pure, and elegant definition of the DLCT which is
fully compatible with the theory of the discrete Fourier transform and
its dual and circulant structure. By self-consistent we mean that the relations between discrete entities should mirror those between continuous entities as much as possible, e.g.\ if the coordinate  multiplication and differentiation operators are dual in the
continuous case, they should also be so in the discrete case. The discrete LCT should be built upon these two operators in the same way that the continuous LCT is, and so forth. By duality we mean that a kind of symmetry between the two domains is exactly satisfied (e.g. coordinate
multiplication in one domain is differentiation in the other,
translation in one domain is phase multiplication in the other, etc.).  All the dual properties of the Fourier transform (such as those in parenthesis above) can be derived from the duality of ${\cal U}$ and ${\cal D}$ \cite{ozaktas01book}, so first and foremost, this duality must be maintained. One of the most important features of our approach
is that our definition maintains this structure by treating both
domains totally symmetrically.

The paper is organized as follows: Section~\ref{pre} reviews the
preliminaries and the definition and important properties of
LCTs. Section~\ref{dlct} describes the theory and derivations for the
proposed DLCT. Theoretical discussions on defining a discrete LCT and the properties of such a definition that need to exist are given in Section~\ref{discussions}. In Section~\ref{results}, numerical examples and comparisons are provided. Lastly, we conclude in Section~\ref{conc}. There is also an Appendix in which we have provided some proofs, necessary fundamental information, justifications and implementation details that are needed for the derivations in Section~\ref{dlct}.

\section{Preliminaries}
\label{pre}
\subsection{Linear Canonical Transform}
LCTs are unitary transforms specified by a $2\times 2$ parameter
matrix $\mathbf{L}$. Because the determinant of $\mathbf{L}$ is
required to be equal to $1$, an LCT can also be uniquely specified by
three independent parameters, often denoted by $\alpha, \beta, \gamma$.
The elements $A, B, C, D$ of the $2\times 2$ matrix and $\alpha,
\beta, \gamma$ are related by:
\begin{equation}
\label{LCTmatrix}
{\mathbf L} = \left[ \begin{array}{cc} A & B \\ C & D \\ \end{array} \right]
= \left[ \begin{array}{cc} \frac{\gamma}{\beta} & \frac{1}{\beta} \\ -\beta+\frac{\alpha \gamma}{\beta} & \frac{\alpha}{\beta} \\ \end{array} \right] = \left[ \begin{array}{cc} \frac{\alpha}{\beta} & \frac{-1}{\beta} \\ \beta-\frac{\alpha \gamma}{\beta} & \frac{\gamma}{\beta} \\ \end{array} \right]^{-1}.
\end{equation}
We can define an LCT through either the parameter set $(A, B, C,
D)$ with the condition that $AD-BC=1$ or the parameter set
$(\alpha, \beta, \gamma)$. In this paper, we restrict ourselves to the
case where the parameters in both sets are all real. The definition of
the LCT as a linear integral transform, using the second set of
parameters, can be written as:
\begin{multline}
\label{lctdef}
{\mathcal C}_{\mathbf L}f(u)= \\ \sqrt{\beta}\,e^{-i\pi/4}\,
\int_{-\infty}^{\infty}
 \exp\left[i\pi(\alpha u^2-2\beta uu'+\gamma u'^2)\right]f(u')\,du'.
\end{multline}
Every triplet $(\alpha, \beta, \gamma)$ corresponds to a different
LCT. We denote the LCT operator using $\mathcal{C}_\mathbf{L}$ where
the subscript $\mathbf{L}$ denotes the $2\times 2$ parameter matrix.

\subsection{Important Properties}

The utility of the parameter set $(A, B, C, D)$ is best appreciated
upon observing the concatenation property: If any two LCTs are
concatenated (applied one after the other), the resulting operation is
also an LCT whose $2\times 2$ matrix is the product of the $2\times 2$
matrices of the two original LCTs. This can be stated as:
\begin{equation} \label{index_add_prop}
\mathcal{C}_\mathbf{L}f(u) = 
\mathcal{C}_\mathbf{L_1} \mathcal{C}_\mathbf{L_2} f(u),
\end{equation}
where $\mathbf{L}=\mathbf{L}_1\mathbf{L}_2$.

An important special case of this property is the reversibility
property. It basically states that the $2\times 2$ matrix for the
inverse of an LCT is again an LCT whose $2\times 2$ matrix is the
matrix inverse of the original LCT:
\begin{equation} \label{rev_prop}
\mathcal{C}_\mathbf{L_2} \mathcal{C}_\mathbf{L_1} f(u) = f(u),
\end{equation}
if $\mathbf{L}_2 = \mathbf{L}^{-1}_1$. 

\subsection{Special Linear Canonical Transforms}
We now give some special transforms and operations, which are all
special cases of LCTs. 

\subsubsection{Scaling}
The parameter matrix for the scaling operation is as follows
\begin{equation}
{\mathbf L_M} = \left[ \begin{array}{cc} M & 0 \\ 0 & \frac{1}{M} \\ \end{array} \right]=\left[ \begin{array}{cc} \frac{1}{M} & 0 \\ 0 & M \\ \end{array} \right]^{-1}.
\end{equation}
Functionally it can be defined in the following way:
\begin{equation}
{\mathcal C}_{{\mathbf L}_{M}}f(u)={\mathcal
M}_{M}f(u)=\sqrt{\frac{1}{M}}\, f\left(\frac{u}{M}\right).
\end{equation}

\subsubsection{Fractional Fourier Transform}
The Fractional Fourier transform (FRT) is the generalized version of
the Fourier transform (FT). It has the following parameter matrix:
\begin{equation}
{\mathbf L}_{{\mathbf F}^{a}_{\rm lc}}=\left[ \begin{array}{cc}
\cos \theta & \sin \theta \\
-\sin \theta & \cos \theta \\
\end{array} \right]=\left[ \begin{array}{cc}
\cos \theta & -\sin \theta \\
\sin \theta & \cos \theta \\
\end{array} \right]^{-1},
\end{equation}
where $\theta = \pi a/2$ and $a$ is the fractional order. When
$a=1$, the FRT reduces to the FT. (It should be noted that there is a
slight difference between the FRT thus defined
($\mathcal{F}^a_{\text{lc}}$) and the more commonly used definition of
the FRT ($\mathcal{F}^a$), \cite{ozaktas01book}.)

The $a$th order fractional Fourier transform $\mathcal{F}^a$ of
the function $f(u)$ may be defined as \cite{ozaktas01book}:
\begin{align}
&\ \mathcal{F} ^af(u) = \int_{-\infty}^{\infty} K_a(u,u') f(u')\,du', \nonumber \\
&K_a(u,u') = A_{\theta} \exp\left[i\pi(u^2\cot\theta-2uu'\csc\theta+u'^2\cot\theta)\right], \nonumber \\ &A_{\theta} =
\frac{\exp(-i\pi\mbox{sgn}(\sin\theta)/4+i\theta/2)}{|\sin\theta|^{1/2}}
\end{align}

\subsubsection{Chirp Multiplication}
The parameter matrix for the chirp multiplication operation is
\begin{equation}
\label{chirpmatrix}
{\mathbf L}_{{\mathbf Q}_q} = \left[ \begin{array}{cc} 1 & 0 \\ -q & 1 \\ \end{array} \right] = \left[ \begin{array}{cc} 1 & 0 \\ q & 1 \\ \end{array} \right]^{-1}.
\end{equation}
The chirp multiplication operation can be expressed as
\begin{equation}
\label{chirpcont}
{\mathcal C}_{{\mathbf Q}_{q}}f(u)={\mathcal Q}_{q}f(u)=\exp (-i\pi qu^{2})f(u).
\end{equation}

Corresponding formulas for chirp convolution may be found in \cite{ozaktas01book}.

\section{Discrete Linear Canonical Transforms}
\label{dlct}

We now present our development of the DLCT based on hyperdifferential
operator theory. Our approach is based on decomposing the LCT into
simpler parts, finding the discrete versions of these parts by using
operator theory, and then multiplying those to obtain the final DLCT
matrix.

Although there are several ways to decompose the LCT \cite{koc08ieee}, here we choose the Iwasawa decomposition since it includes a greater number of special LCTs than other decompositions, providing the opportunity to discuss their hyperdifferential forms. The method of using hyperdifferential operators outlined here can also be applied to other decompositions.

\subsection{The Iwasawa Decomposition}

The linear canonical transform (LCT) operator $\mathcal{C}_\mathbf{L}$
can be expressed as combinations of other simpler operators in many
ways. Using scaling $\mathcal{M}_M$, chirp multiplication
$\mathcal{Q}_q$ and fractional Fourier $\mathcal{F}^a$ operators, it
is possible to construct any linear canonical transform. The Iwasawa
decomposition we will employ, breaks down an arbitrary LCT into a
fractional Fourier transform followed by scaling followed by chirp
multiplication, and can be written in operator notation as follows
\cite{ozaktas16lctbook}:
\begin{equation} \label{decomposition1}
\mathcal{C}_\mathbf{L} = \mathcal{Q}_q \mathcal{M}_M \mathcal{F}_{\text{lc}}^a ,
\end{equation}

When each operator is characterized by their $2 \times 2$ LCT
parameter matrix, the decomposition looks like
\begin{align}
{\mathbf L} & = \left[ \begin{array}{cc} A & B \\ C & D \\ \end{array} \right] = \left[ \begin{array}{cc} \frac{\gamma}{\beta} & \frac{1}{\beta} \\ -\beta+\frac{\alpha \gamma}{\beta} & \frac{\alpha}{\beta} \\ \end{array} \right] \nonumber\\
& = \left[ \begin{array}{cc} 1 & 0 \\ -q & 1 \\ \end{array}\right]
    \left[ \begin{array}{cc} M & 0 \\ 0 & 1/M \\ \end{array}\right]
    \left[ \begin{array}{cc} \cos a\pi/2 & \sin a\pi/2 \\
    -\sin a\pi/2 & \cos a\pi/2 \\ \end{array}\right]   \label{decomp1matrix} 
\end{align}
where $a$, $q$, $M$ must be chosen as:
\begin{align}
&M = \left \{ \begin{array}{rl} \sqrt{1+\gamma^{2}}/\beta, &
\gamma
\geq 0, \\
-\sqrt{1+\gamma^{2}}/\beta, & \gamma  <0, \end{array}
\right. \label{m} \\
&q = \frac{\gamma \beta^2}{1+\gamma^2}-\alpha, \label{q} \\
&a = \frac{2}{\pi} \arccot \gamma. \label{a}
\end{align}

This decomposition can break down any arbitrary linear canonical
transform into a cascade of elementary operations. Our approach will
be to find the $N\times N$ discrete transform matrix for each of these
three operations and multiply them to obtain the discrete LCT matrix.

\subsection{The Hyperdifferential Forms}

The term hyperdifferential refers to having differential operators in an exponent. In the LCT context, we only have second order coordinate multiplication and differentiation operators in the exponent. Operators representing an arbitrary LCT or all of its special cases can be generated by exponentiating these second order operators and these constitute the hyperdifferential forms of these transforms. There is correspondence among the integral transforms, hyperdifferential operators and the 2x2 parameter matrices that are given in the preliminaries section. An LCT can be represented by any one of these mathematical objects. More details can be found in \cite{wolf79book}.

It is well established that the chirp multiplication operator
$\mathcal{Q}_q$, the scaling operator $\mathcal{M}_M$, and the
fractional Fourier transform operator $\mathcal{F}^a_{\text{lc}}$ 
can all be written in hyperdifferential forms as follows:
\cite{wolf79book,ozaktas01book}: 
\begin{equation} \label{cm_op}
\mathcal{Q}_q=\exp{\left(-i2\pi q \,
\frac{\mathcal{U}^2}{2}\right)},
\end{equation}
\begin{equation} \label{scaling_op}
\mathcal{M}_M=\exp{\left(-i2\pi \ln{(M)} \,
\frac{\mathcal{UD}+\mathcal{DU}}{2}\right)},
\end{equation}
\begin{equation} \label{fracF_op}
\mathcal{F}^a_{\text{lc}}=\exp{\left(-ia\pi^2 \,
\frac{\mathcal{U}^2+\mathcal{D}^2}{2}\right)},
\end{equation}
where $\mathcal{U}$ and $\mathcal{D}$ are the coordinate
multiplication and differentiation operators,
respectively. We see that all three of the operators we are working
with can be expressed in terms of these two building blocks, whose
continuous manifestations are:
\begin{eqnarray}
{\cal U} f(u) = uf(u) \\
{\cal D} f(u) = \frac{1}{i2\pi} \frac{df(u)}{du},
\end{eqnarray}
where the $(i2\pi)^{-1}$ is included so that ${\cal U}$ and ${\cal D}$
are precisely Fourier duals (the effect of either in one domain is its
dual in the Fourier domain). This duality can be expressed as follows: 
\begin{equation} \label{U_fdf}
\mathcal{U}=\mathcal{FDF}^{-1}.
\end{equation}

\subsection{The Discrete Linear Canonical Transform}

Our approach is based on requiring that, to the extent possible, all
the discrete entities we define observe the same structural
relationships as they do in abstract operator form. We want a discrete definition that is as analogous to the continuous definition as possible. To ensure this, we define
the discrete LCT and its special cases as the discrete manifestations of
Eq.~\ref{decomposition1}, Eq.~\ref{cm_op}, Eq.~\ref{scaling_op} and
Eq.~\ref{fracF_op}, with the
abstract operators being replaced by matrix operators. This can be written as follows:
\begin{equation} \label{decomp1matrix2}
\mathbf{C}_\mathbf{L} = \mathbf{Q}_q \mathbf{M}_M \mathbf{F}_{\text{lc}}^a.
\end{equation}
\begin{equation} \label{cm_mat}
\mathbf{Q}_q=\exp{\left(-i2\pi q \,
\frac{\mathbf{U}^2}{2}\right)}.
\end{equation}
\begin{equation} \label{scaling_mat}
\mathbf{M}_M=\exp{\left(-i2\pi \ln{(M)} \,
\frac{\mathbf{UD}+\mathbf{DU}}{2}\right)}.
\end{equation}
\begin{equation} \label{fracF_mat}
\mathbf{F}^a_{\text{lc}}=\exp{\left(-ia\pi^2 \,
\frac{\mathbf{U}^2+\mathbf{D}^2}{2}\right)}.
\end{equation}
Note that $\exp()$ in the above equations are matrix
exponentials and how they are computed is discussed in
Appendix~\ref{expm}. Thus the discrete LCT matrix is given by
\begin{eqnarray} \label{lct_full}
\mathbf{C}_\mathbf{L} = \exp{\left(-i2\pi q \,
\frac{\mathbf{U}^2}{2}\right)} \times  \nonumber \\
\exp{\left(-i2\pi \ln{(M)} \,
\frac{\mathbf{UD}+\mathbf{DU}}{2}\right)}\exp{\left(-ia\pi^2 \,
\frac{\mathbf{U}^2+\mathbf{D}^2}{2}\right)}.
\end{eqnarray}
The discrete LCT matrix is defined as the product of the FRT, scaling,
and chirp multiplication matrices, all of which are defined in terms
of the $\mathbf{U}$ and $\mathbf{D}$ matrices. To get the DLCT of a 
function of a discrete variable, we just need to write it as a column
vector and multiply it with the DLCT matrix
$\mathbf{C}_\mathbf{L}$.

Thus it is seen that all rests on the differentiation and coordinate
multiplication matrices $\mathbf{D}$ and $\mathbf{U}$ and computation
of the matrix exponentials in Eq.~\ref{lct_full}.
Thus, we move on to how to obtain the $\mathbf{U}$ and $\mathbf{D}$
matrices. 

For signals of discrete variables, the closest thing to differentiation is finite differencing. Consider the following definition:
\begin{equation} \label{D_defn}
\tilde{\mathcal{D}}_hf(u)=\frac{1}{i2\pi}\frac{f(u+h/2)-f(u-h/2)}{h}.
\end{equation}
If $h\rightarrow 0$, then $\tilde{\mathcal{D}}_h \rightarrow \mathcal{D}$, since in this case the right-hand side approaches $(i2\pi)^{-1}df(u)/du$. Therefore, $\tilde{\mathcal{D}}_h$ can be interpreted as a finite difference operator.

Now, using $f(u+h)=\exp(i2\pi h\mathcal{D})f(u)$, which is another established result in operator theory \cite{wolf79book,ozaktas01book}, we express Eq.~\ref{D_defn} in hyperdifferential form:
\begin{align} \label{D_hyp}
\tilde{\mathcal{D}}_h & =\frac{1}{i2\pi}
\frac{e^{i\pi h\mathcal{D}}-e^{-i\pi h\mathcal{D}}}{h} \nonumber \\
& = \frac{1}{i2\pi} \frac{2i\sin(\pi h\mathcal{D})}{h}
={\rm sinc}(h\mathcal{D}) \;\mathcal{D}.
\end{align}
Note that if we let $h\rightarrow 0$ in the last equation and take the limit, we can verify that $\tilde{\mathcal{D}}_h \rightarrow \mathcal{D}$ from here as well. 

Now, we turn our attention to the task of defining $\tilde{\mathcal{U}}_h$. It is tempting to define the discrete version of the coordinate multiplication matrix by simply forming a diagonal matrix with the diagonal entries being equal to the coordinate values. However, upon closer inspection we have decided that this could not be taken for granted. In order to obtain the most self-consistent formulation possible, we must be sure to maintain the structural symmetry between $\mathcal{U}$ and $\mathcal{D}$ in all their manifestations. Therefore, we choose to define $\tilde{\mathcal{U}}_h$ such that it is related to $\mathcal{U}$, in exactly the same way as $\tilde{\mathcal{D}}_h$ is related to $\mathcal{D}$:
\begin{equation}
\tilde{\mathcal{U}}_h = {\rm sinc}(h\mathcal{U}) \;\mathcal{U},
\end{equation}
from which we can observe that as $h\rightarrow 0$, we have $\tilde{\mathcal{U}}_h \rightarrow \mathcal{U}$, as should be. However, beyond that, it is also possible to show that, $\tilde{\mathcal{U}}_h$, when defined like this, satisfies the same duality expression Eq.~\ref{U_fdf} satisfied by $\mathcal{U}$ and $\mathcal{D}$:
\begin{equation}
\tilde{\mathcal{U}}_h = \mathcal{F} \tilde{\mathcal{D}}_h \mathcal{F}^{-1}.
\end{equation}
To see this, substitute $\tilde{\mathcal{D}}_h$ in this equation:
\begin{align}
\label{sincUU}
\tilde{\mathcal{U}}_h & = \mathcal{F} \left(\frac{1}{i2\pi} \frac{2i\sin(\pi h\mathcal{D})}{h}\right)  \mathcal{F}^{-1} \nonumber \\
& = \frac{1}{i2\pi} \frac{2i\sin(\pi h\mathcal{U})}{h}
={\rm sinc}(h\mathcal{U}) \mathcal{U}.
\end{align}
When acting on a continuous signal $f(u)$, the operator $\mathcal{U}$ becomes
\begin{equation} \label{U_final_op}
\tilde{\mathcal{U}}_hf(u)=\frac{1}{\pi} \frac{\sin(\pi hu)}{h} f(u).
\end{equation}
We observe that the effect is not merely multiplying with the coordinate variable. Had we defined $\tilde{\mathcal{U}}_h$ such that it corresponds to multiplication with the coordinate variable, we would have destroyed the symmetry and duality between $\mathcal{U}$ and $\mathcal{D}$ in passing to the discrete world.

Now, by sampling Eq.~\ref{U_final_op}, we can obtain the matrix operator to act on finite discrete signals. The sample points will be taken as $u=nh$ to finally yield the $\mathbf{U}$ matrix defined as:
\begin{equation}\label{Umatrix}
U_{mn}= \begin{cases}
\frac{\sqrt{N}\,\;}{\pi} \sin \left(\frac{\pi}{N}n \right), & \text{for } m=n \\
0, & \text{for } m \neq n
\end{cases}.
\end{equation}
As always, the value of $N$ should be determined based on the time/space and frequency extent of the signal, along with the required accuracy \cite{koc08ieee,gulcu18choiceQ,ozcelikkale12finitebits,ozcelikkale13randomfields}. Further detail is provided in Section~\ref{subsectionsampling}.

The matrix $\mathbf{D}$, on the other hand, can be calculated in terms of
$\mathbf{U}$ by using the discrete version of the duality relation
given in Eq.~\ref{U_fdf}:
\begin{equation} \label{D_fuf}
\mathbf{D}=\mathbf{F}^{-1}\mathbf{UF},
\end{equation}
in which $\mathbf{F}$ is the matrix representing the unitary discrete
Fourier transform (DFT) matrix. The elements $F_{mn}$ of the
$N$-point unitary DFT matrix $\mathbf{F}$ can be written in terms of
$W_N = \exp(-j 2\pi/N)$ as follows:
\begin{equation*} \label{DFT}
F_{mn}= \frac{1}{\sqrt{N}\,} W_N^{mn}.
\end{equation*}

When all is put together, the LCT of a signal $x[n]$ of length $N$, represented by the column vector $\mathbf{x}$, is
then computed by $\mathbf{C}_\mathbf{L}\mathbf{x}$, yielding an $N\times 1$
output. Further details of the development of the $\mathbf{U}$ and
$\mathbf{D}$ matrices and their applications may be found in
\cite{koc18scaling}, which together with the present work, not only
establish a formulation of these operators that is fully consistent
with the theory of the DFT and its circulant structure, but also pave
the way for the utilization of operator theory in deriving other more
sophisticated discrete operations. We believe these works
are the first to apply operator theory in defining discrete transforms.

\subsection{Unitarity of the Discrete Linear Canonical Transform}
\label{proofunitarity}

One of the most essential properties of the kind of discrete
transforms we are working with is unitarity. This leads to Parseval
type relationships and manifests itself as energy or power
conservation in physical applications.

Here we prove that the proposed DLCT definition is unitary by showing
that the matrix $\mathbf{C}_\mathbf{L}$ given in
Eq.~\ref{decomp1matrix2} and more explicitly in Eq.~\ref{lct_full} is
unitary. 

\begin{theorem}
\label{theorem1}
The discrete LCT defined in Eq.~\ref{lct_full} is unitary, with $M, q, a$ chosen according to Eqs.~\ref{m}, \ref{q}, \ref{a}, and ${\bf U}$ and ${\bf D}$ defined according to Eqs.~\ref{Umatrix} and \ref{D_fuf}.
\end{theorem}

Before proceeding with the proof, we first recall some fundamental
definitions: A matrix $\mathbf{A}$ is said to be Hermitian when
$\mathbf{A}=\mathbf{A^H}$ holds, where $\mathbf{A^H}$ denotes the
conjugate transpose of $\mathbf{A}$, and is said to be unitary when
$\mathbf{A}^{-1}=\mathbf{A^H}$. Since $\mathbf{C}_\mathbf{L}$ is
defined as the product of three matrices, showing that each of
them is unitary will suffice to show that $\mathbf{C}_\mathbf{L}$ is
unitary. $\mathbf{U}$ and $\mathbf{D}$ are the fundamental matrices
that give rise to those three components. We will first show that
these matrices are Hermitian. From that it will follow that the three
multiplied matrices are all unitary.

\begin{theorem}
\label{theorem2}
The matrices ${\bf U}$ and ${\bf D}$ are Hermitian and the matrices defined in Eqs.~\ref{cm_mat}, \ref{scaling_mat}, \ref{fracF_mat} are unitary.
\end{theorem}

Theorem \ref{theorem2} is proved in the Appendix~\ref{appendix_unitarity} from which Theorem \ref{theorem1} follows.

\subsection{Discretization, Sampling and Indexing}
\label{subsectionsampling}

We introduce discretization by replacing the continuous derivative with a finite difference, such that, as the finite interval goes to zero, it approaches the continuous derivative. Remembering that exponentiation etc.\ can be expressed as power series, the full LCT development is then based on the following operations on this finite difference operation: inversion, fractional and ordinary Fourier transformation, repeated application, multiplication with a scalar and addition. Now, as the finite difference goes to a derivative, similar will hold for its repeated applications, as well as scalar multiplied and added versions. Likewise, we know that the DFT approximates the continuous Fourier transform more and more closely as the sampling interval is reduced, so if this operation is in succession with finite differencing, the resulting limit will be the succession of Fourier transformation and continuous differentiation. Similar applies to fractional Fourier transformation, of which inversion is a special case.

In this paper we deal with finite-length signals of a discrete
(integer) variable. (We could equivalently think of them as being
defined on a circulant domain, which would not make a difference in
our arguments.) The length of our signal vectors will be denoted by
$N$. When $N$ is even, they will be defined on the interval of
integers $[-\frac{N}{2},\frac{N}{2}-1]$, and when $N$ is odd, they
will be defined on the interval of integers
$[-\frac{N-1}{2},\frac{N-1}{2}]$. We will also consider an
alternative, less-common approach based on the device of using ``half
integers.'' In this approach, the domain is defined as the interval of
unit-spaced half integers $[-\frac{N}{2}+0.5,\frac{N}{2}-1+0.5]$ for
even $N$ and $[-\frac{N-1}{2}-0.5,\frac{N-1}{2}-0.5]$ for odd
$N$. Although not very usual, there is nothing unnatural about this
way of indexing signals of a discrete variable; it is merely a
particular way of bookkeeping. Note that the indices are still spaced
by unity, and there is merely a shift by $0.5$ with the purpose of
making the interval symmetrical around the origin when $N$ is even
(with the consequence that symmetry is lost when $N$ is odd). A few
examples of works considering this way of indexing are
\cite{grunbaum82centeredDFT,clary03shiftedFourier,mugler11centeredDFT,vargasrubio05centered}. Consistent
with this literature, we will refer to the former approach as the
\textit{ordinary\/} DFT and refer to the latter one, in which we use
"half integers", as the \textit{centered\/} DFT. The DLCT derivation procedure we presented has
been carefully written in a manner that it is consistent with both
approaches. Readers interested in further details on this issue may
refer to \cite{koc18scaling}.

How the number of samples $N$ should be chosen will be determined by factors such as the temporal or spatial extent of the signal, the frequency extent of the signal and therefore the time- or space-bandwidth product. It will also depend on the precision with which the results need to be computed in that application. The choice of $N$ is exogenous to our method. Nevertheless, for completeness, let us elaborate on how the number of samples $N$ is chosen. If the temporal or spatial extent is $\Delta x$ and the double-sided frequency extent is $\Delta\nu$, then we should be sampling with an interval of $1/\Delta\nu$, which means $\Delta x/(1/\Delta\nu)=\Delta x \Delta\nu$ samples. We call this number of samples $N$, the time- or space- bandwidth product. If appropriate normalization as described in \cite{koc08ieee} is applied so that the time/space extent and the frequency extent are made equal in a dimensionless space, it follows that we should sample over an extent $\sqrt{N}\,$ with sampling interval $h=1/\sqrt{N}\,$. Thus as we increase $N$, we will be making $h$ smaller and smaller. Consequently, the finite difference operator in Eq.~\ref{D_defn} approaches a continuous derivative and the finite coordinate multiplication operator will approach the continuous coordinate multiplication operator. The matrix in Eq.~\ref{Umatrix} will approach $U_{mn} = n/\sqrt{N}\,$, corresponding to samples of continuous coordinate multiplication. Since all our operators, including the LCT, are defined in terms of coordinate multiplication and differentiation through smooth exponential functions, they will all approach their continuous counterparts.

\section{Discussions}
\label{discussions}

Continuous unitary LCTs represented by the parameter matrices
$\mathbf{L}$ form the real symplectic group $Sp(2,R)$ with three
independent parameters \cite{moshinsky71lct}. The desirable properties of a discrete LCT mirror those of the
continuous LCT: unitarity, preservation of group structure as
expressed by the concatenation property (and its special case
reversibility), reduction to important special cases and inverses of special cases, and some satisfactory approximation of the continuous
transform. However, a theorem from group theory
\cite{wolf16lctchapter,knapp01bookgroup} precludes realization of this
ideal: It is theoretically impossible to discretize all LCTs with
a finite number of samples such that they are both unitary and they
preserve the group structure
\cite{wolf16lctchapter,knapp01bookgroup}. More on the
group-theoretical properties of LCTs can be found in
\cite{wolf79book,wolf16lctchapter,ozaktas01book}. 

That said, no unitary DLCT definition can exhibit exact
concatenation/reversibility properties. However, if the proposed definition
is to have practical use, we can expect that these properties are at
least approximately satisfied. In Section \ref{proofunitarity}, we
theoretically proved that our proposed DLCT is unitary, so that it
cannot exactly satisfy the concatenation/reversibility property.
Therefore, in the next section, we will numerically show that the
concatenation and reversibility properties are satisfied with a
reasonable accuracy. We will also show that, regardless of
concatenation, the discrete transform provides a reasonable
approximation to the continuous LCT. Before moving on, it needs to be noted that our definition, by construction, reduces to the identity, Fourier and fractional Fourier transforms, chirp multiplication, and magnification (scaling). This result can be trivially obtained by substituting the combination of values leading to the special cases for the parameters $a$, $M$, and $q$ in Eq.~\ref{lct_full}.

\section{Numerical Results and Comparisons}
\label{results}

We will numerically explore three different aspects of the proposed
DLCT definition: (i) approximation of the continuous LCT, (ii)
concatenation of multiple transforms, and (ii) reversibility. We will
carry out numerical tests regarding these aspects of the proposed DLCT
definition.

As the example input functions, the discretized versions of the
chirped pulse function $ \exp (-\pi u^2 - i \pi u^2)$, denoted F1, the trapezoidal function $1.5{\rm tri} (u/3) - 0.5{\rm tri}(u)$, denoted F2
(${\rm tri}(u) = {\rm rect}(u)* {\rm rect}(u)$), rectangular pulse function
${\rm rect}(u)$, denoted F3, and the damped sine function $\exp(-2 | u |) \sin(3\pi u) $, denoted F4, are used. The number of samples $N$ are taken as 256 and 1024 for two sets of numerical simulations. Four
transforms, denoted by T1, T2, T3, and T4, are considered, with parameters $ (\alpha,
\beta, \gamma) = (-3,-2,-1) $, $(-0.8,3,1)$, $(-1.8,-1.75,-1.3)$, and $(0.3,-1.6,-0.9)$, respectively. The LCTs T1, T2, T3 and T4 of the functions 
F1, F2, F3 and F4 have been computed both by the presented DLCT and by a
highly inefficient brute force numerical approach which is taken as a
reference. Throughout our numerical comparisons we use percentage mean squared error (MSE) as the performance metric. It is defined as the energy of the difference normalized by the energy of the reference, expressed as a percentage.

\subsection{Approximation of the Continuous LCT}

\begin{figure*}[h!]
          \subfigure[Real part of T1 of F1]{
          \label{fig:5-2}
            \centering
          \includegraphics[width=0.48\textwidth]{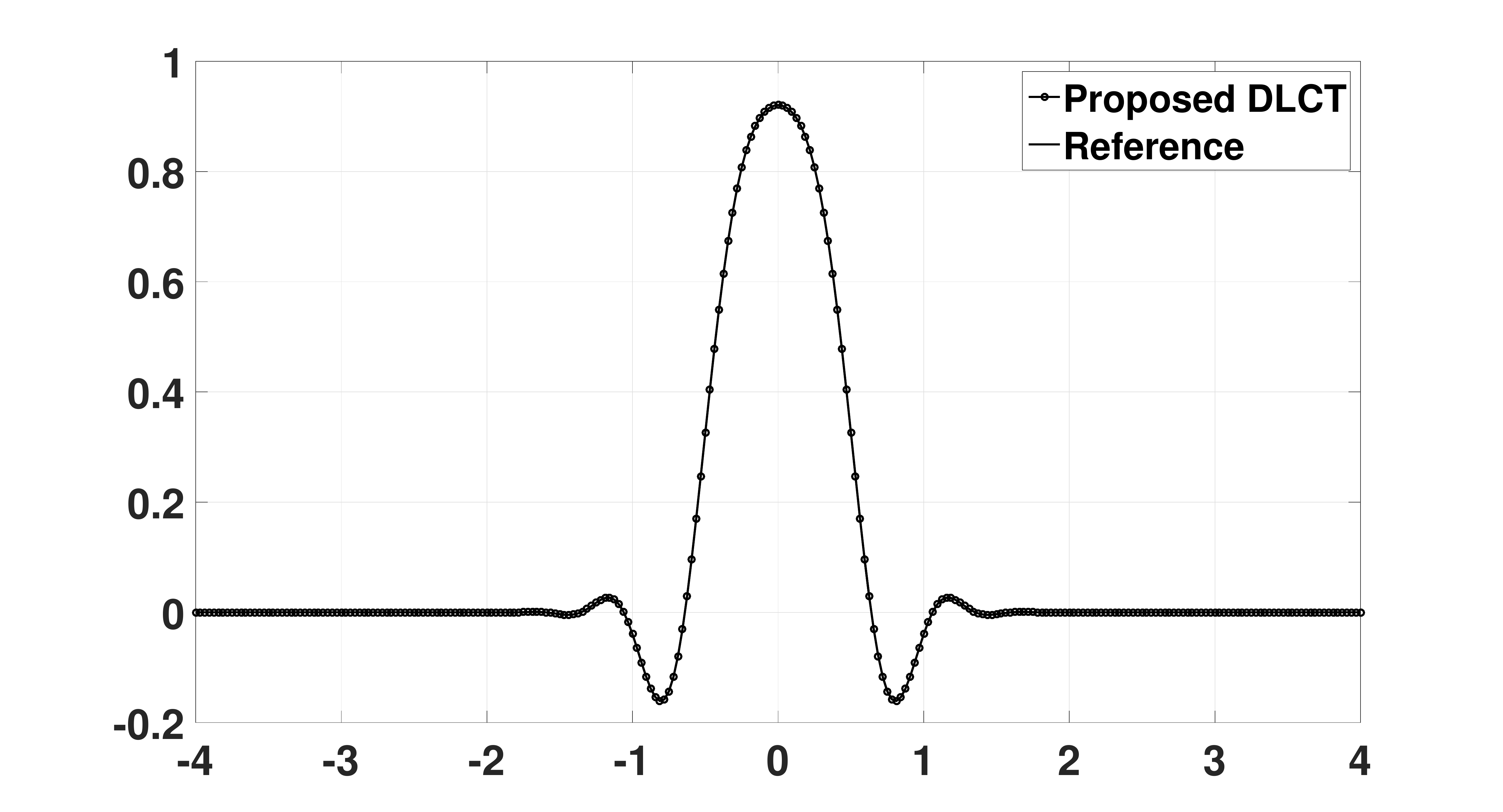}}
           \centering
          \subfigure[Imaginary part of T1 of F1]{
          \label{fig:5-3}
           \centering
           \includegraphics[width=0.48\textwidth]{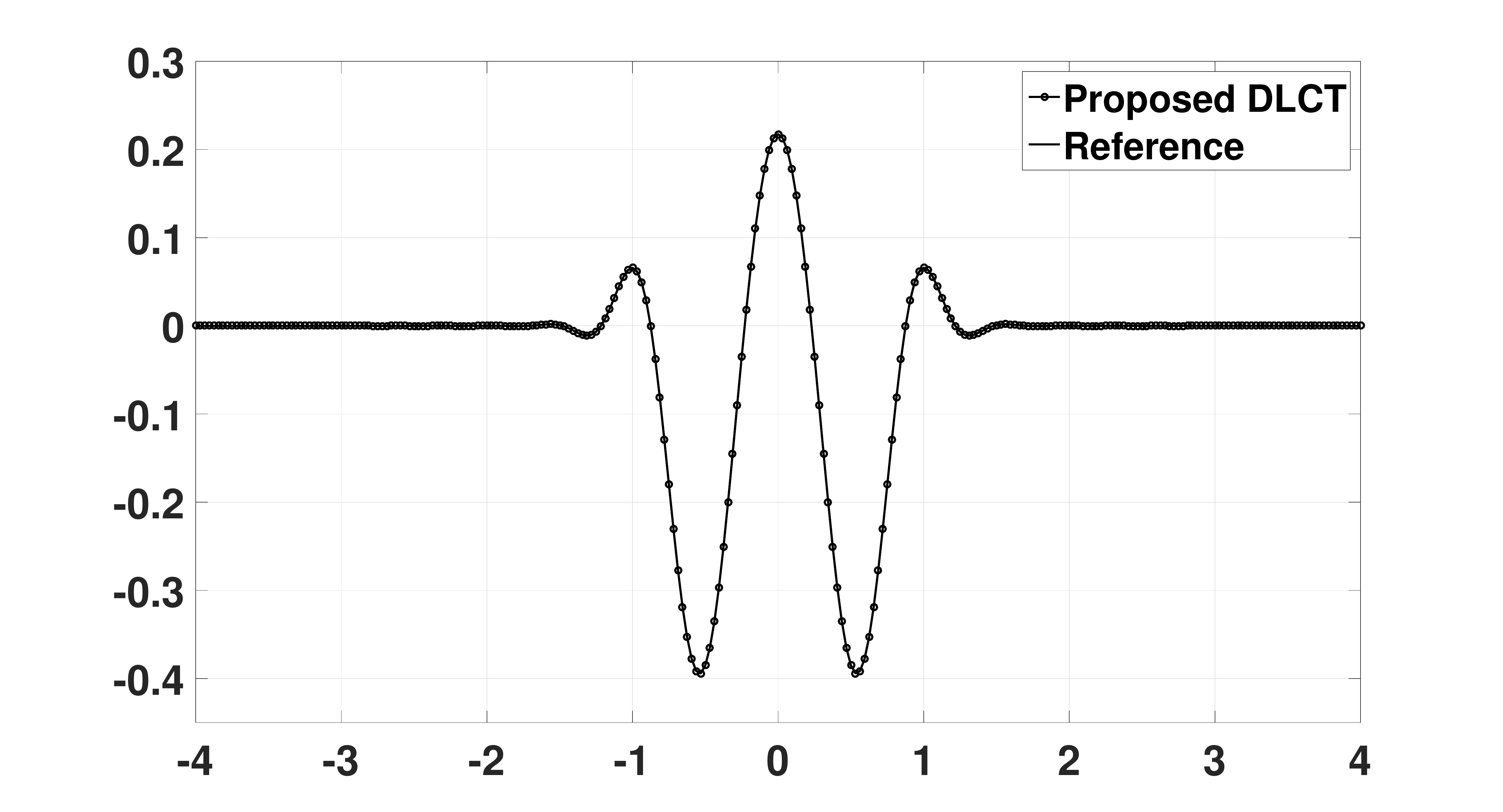}}

      	  \subfigure[Real part of T2 of F2]{
          \label{fig:5-4}
           \centering
           \includegraphics[width=0.48\textwidth]{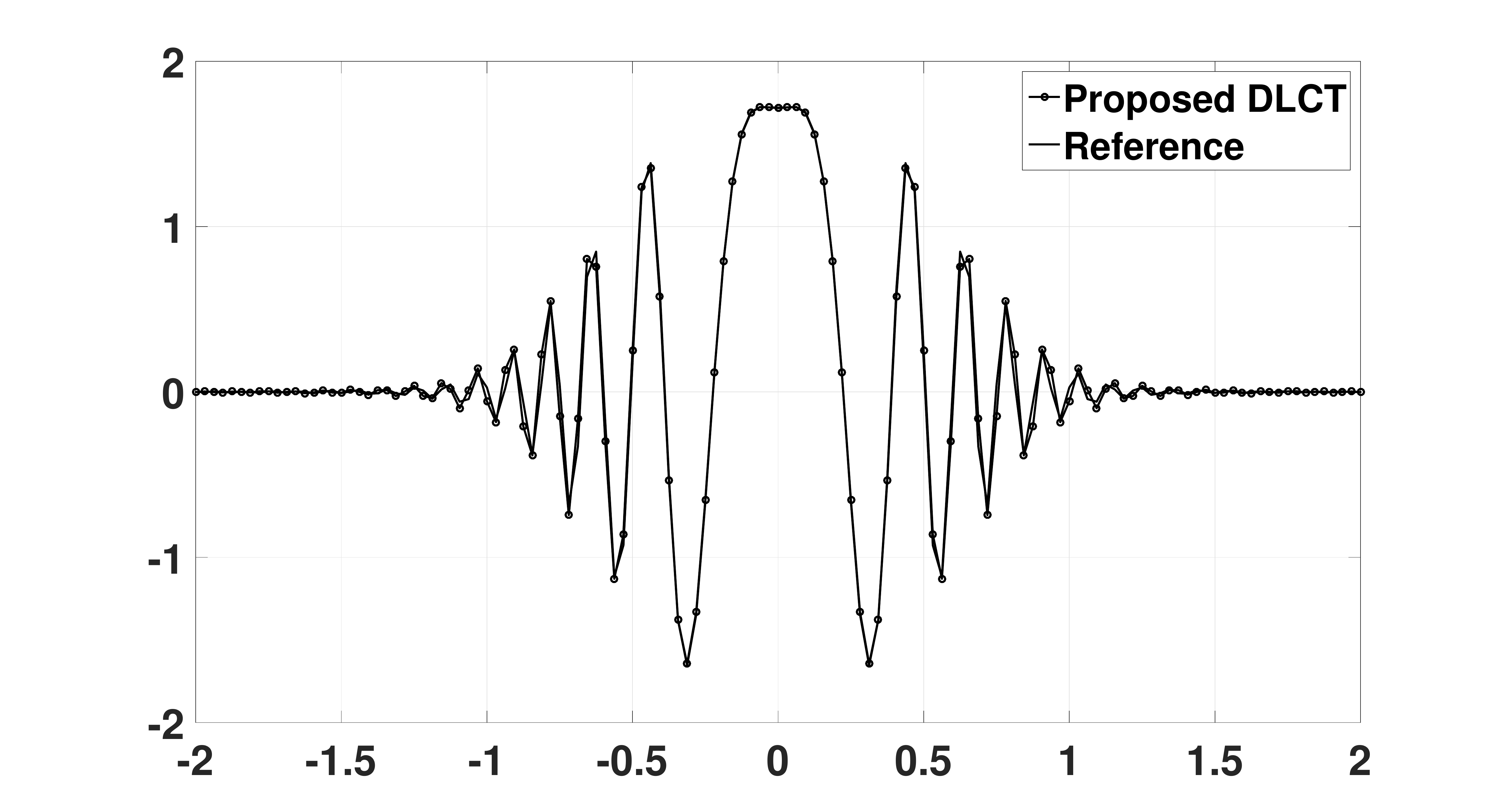}}
           \subfigure[Imaginary part of T2 of F2]{
          \label{fig:5-4}
           \centering
           \includegraphics[width=0.48\textwidth]{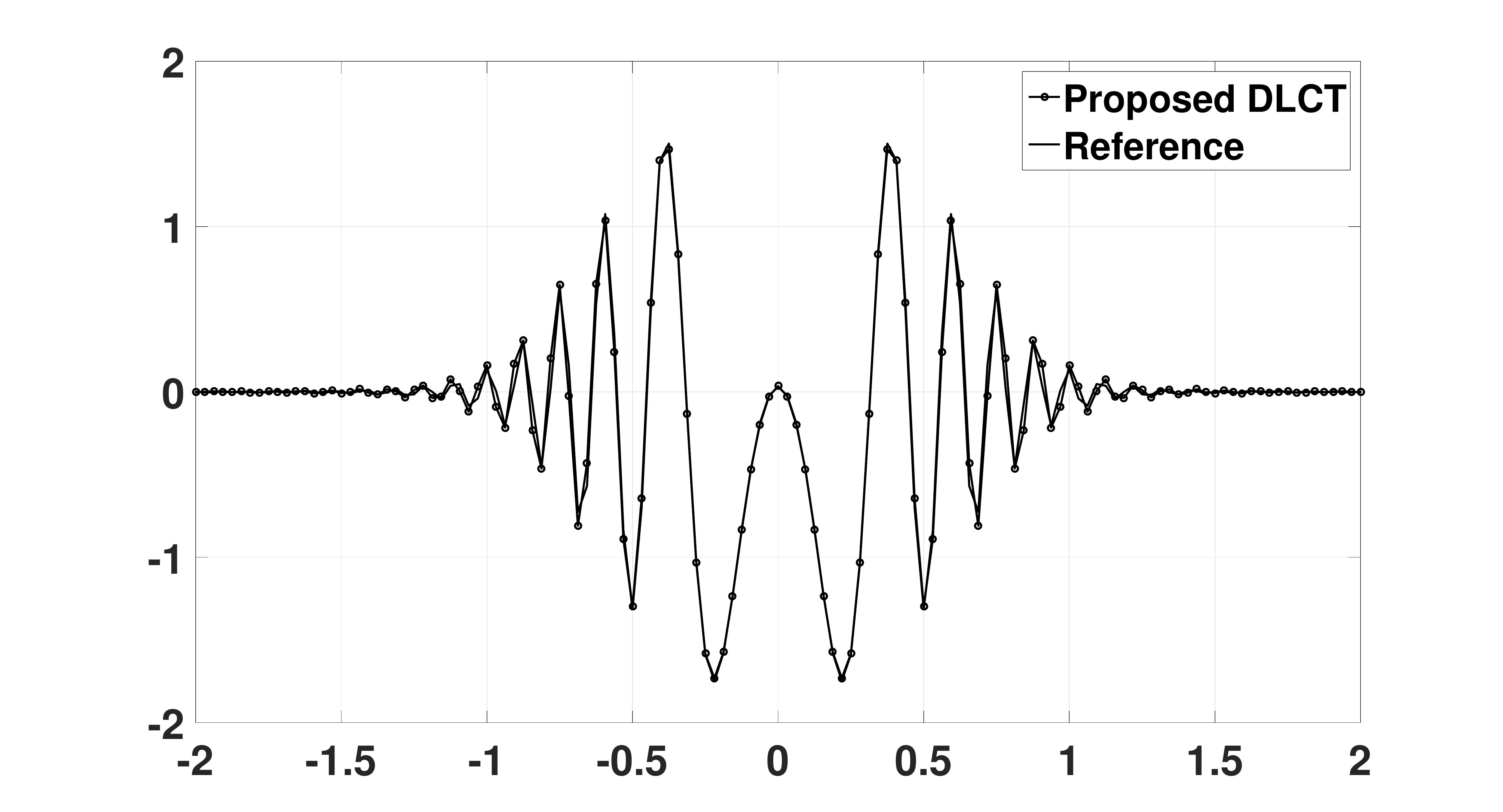}}
           \subfigure[Real part of T3 of F3]{
          \label{fig:5-4}
           \centering
           \includegraphics[width=0.48\textwidth]{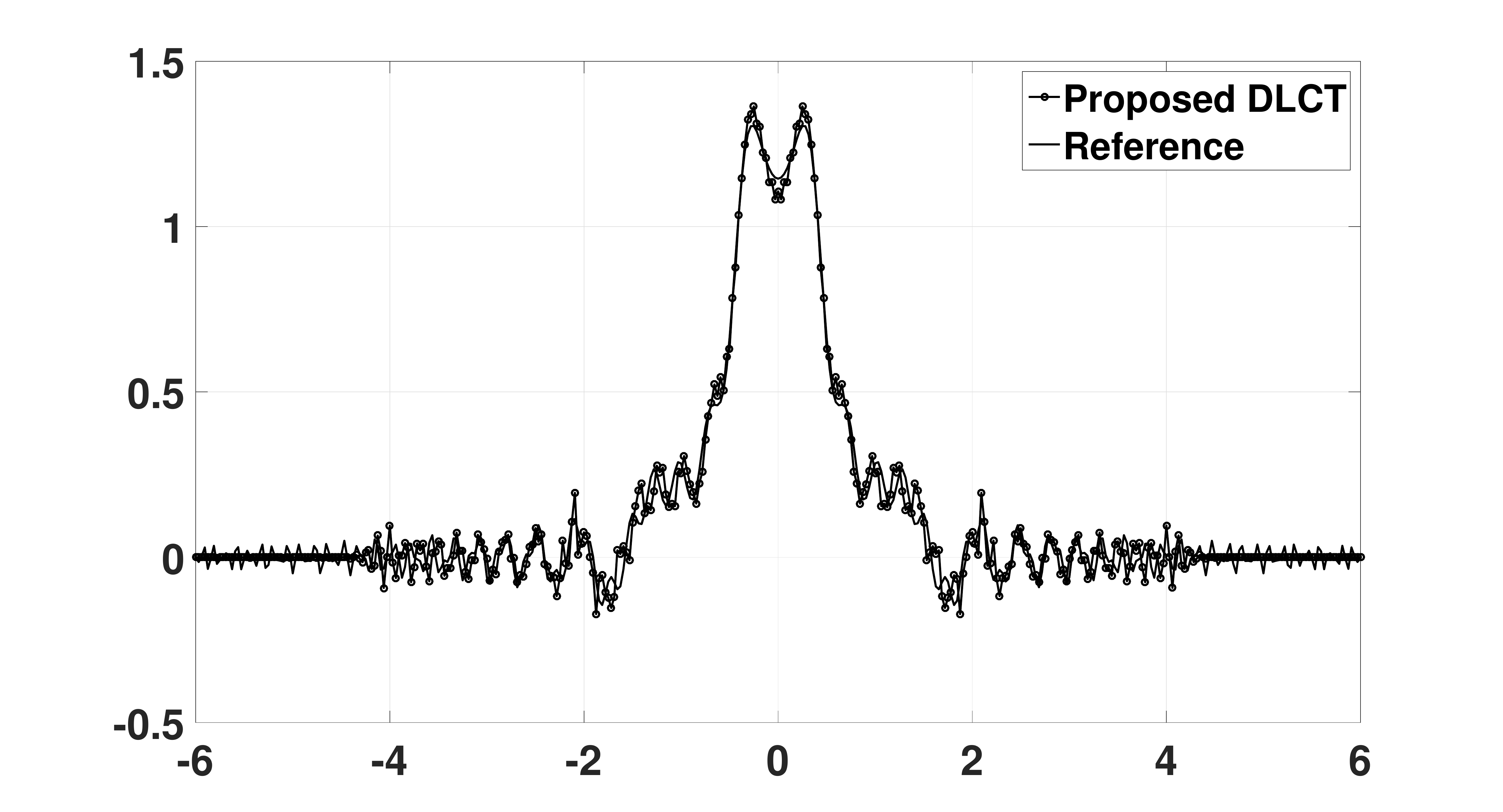}}
          \subfigure[Imaginary part of T3 of F3]{
          \label{fig:5-2}
          \centering
          \includegraphics[width=0.48\textwidth]{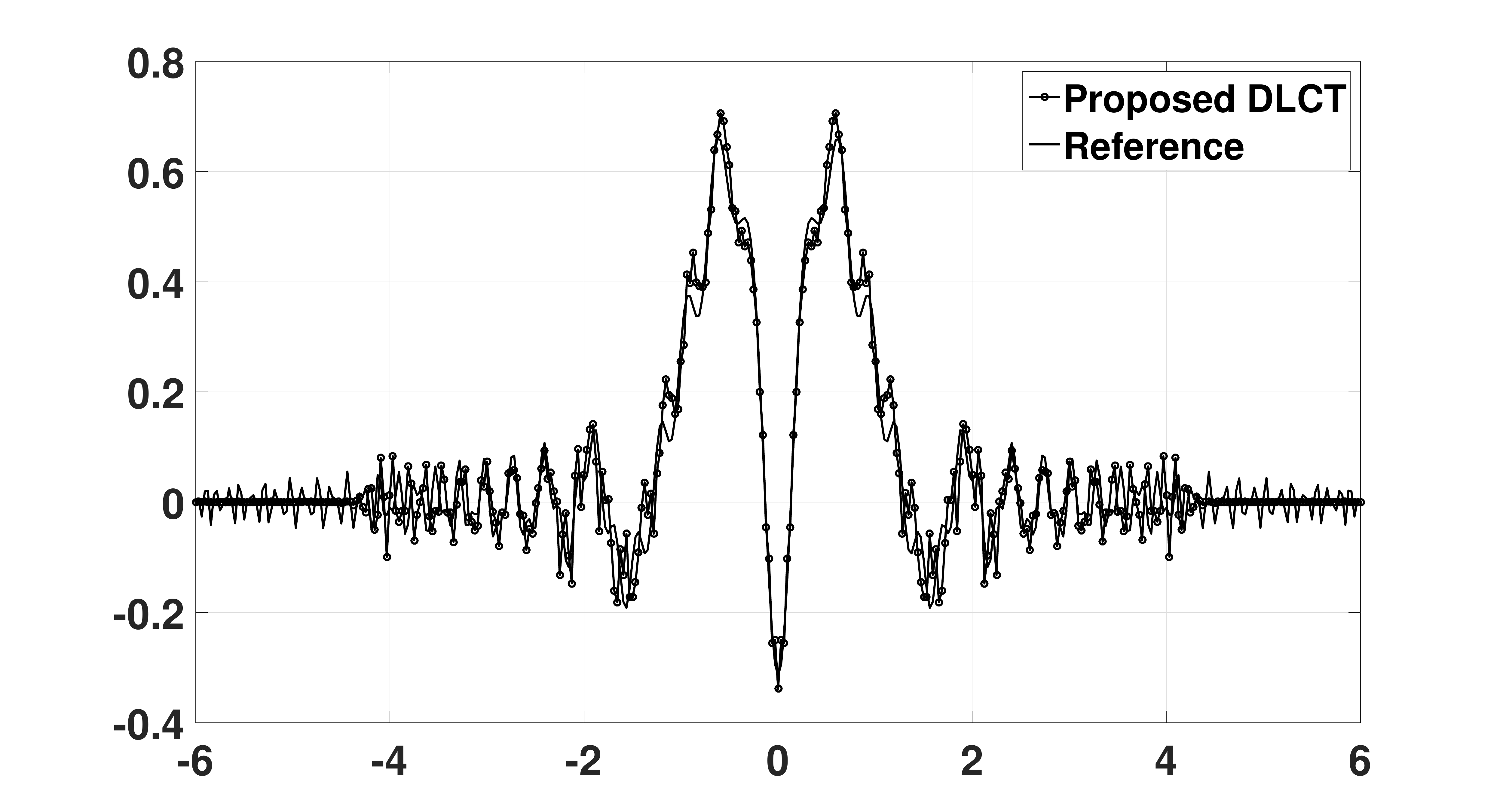}}
                    \subfigure[Real part of T4 of F4]{
          \label{fig:5-2}
          \centering
          \includegraphics[width=0.48\textwidth]{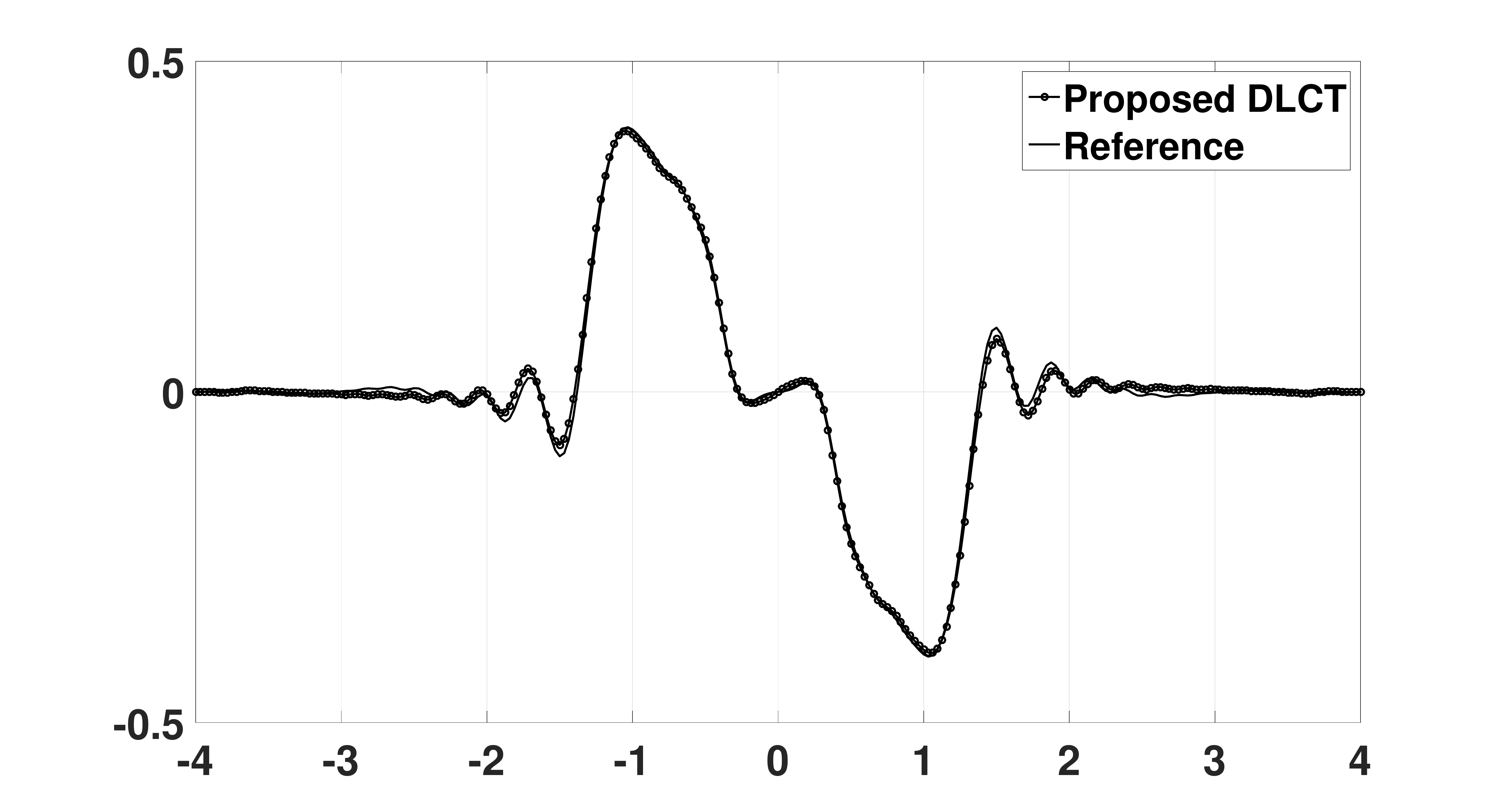}}
                    \subfigure[Imaginary part of T4 of F4]{
          \label{fig:5-2}
          \centering
          \includegraphics[width=0.48\textwidth]{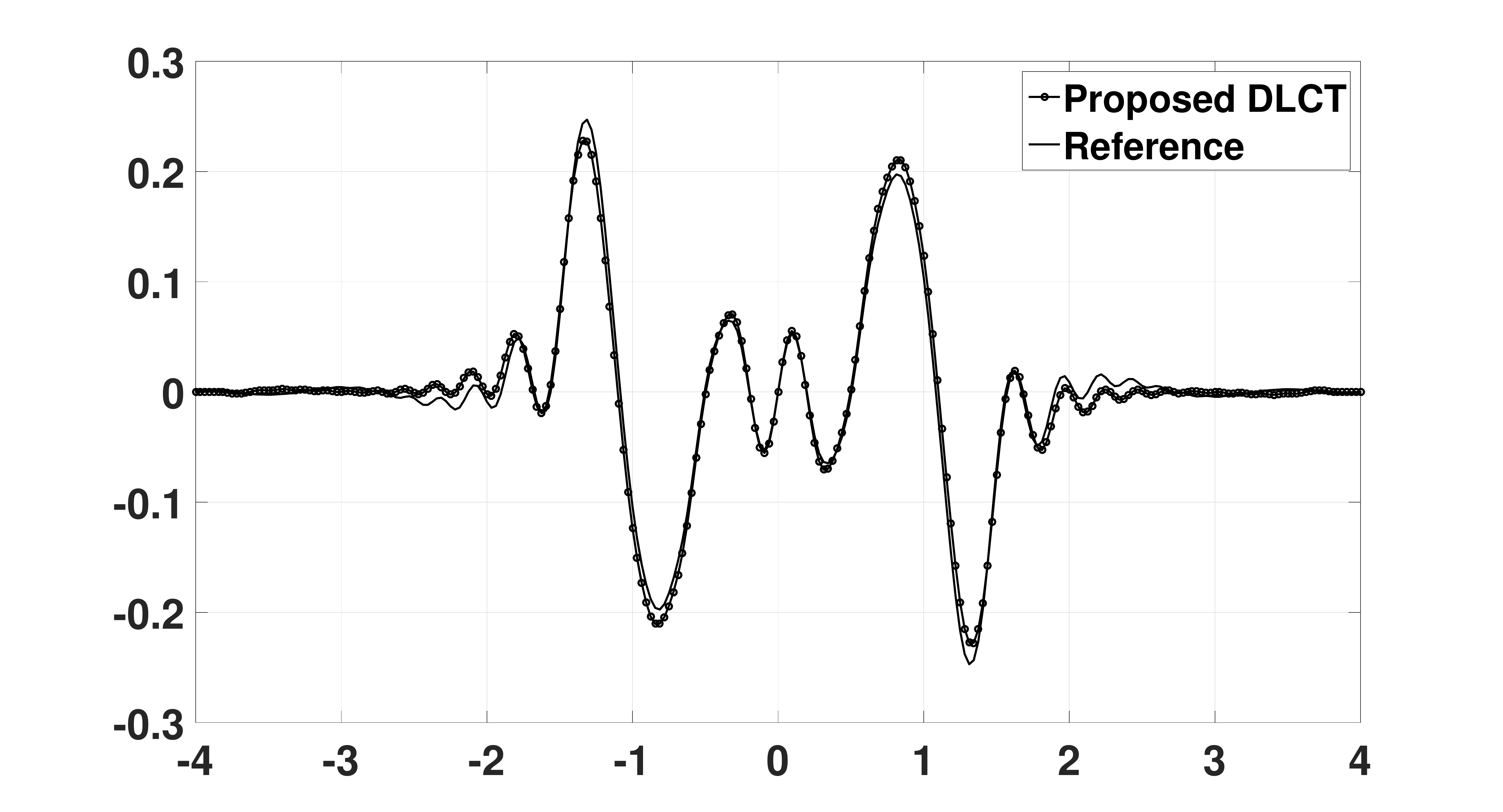}}
       \centering 
  
\caption{Comparison of the proposed DLCT of functions with the reference.}

\label{plots}
\end{figure*}

In this subsection, we focus on how well our method approximates
the continuous LCT. The ``true'' continuous LCT of the original
function is obtained by highly inefficient brute force numerical
integration of the continuous LCT. The resulting percentage MSE scores, for both
\textit{ordinary} and \textit{centered} sampling schemes, turn out to be giving
very similar results, are tabulated in 
Table~\ref{mse_scores_table_ord}. Plots for some examples for the resulting DLCTs (T1 of F1, T2 of F2, T3 of F3 and T4 of F4) and the corresponding references obtained by the brute force numerical method have been presented for both real and imaginary parts of the signals in Fig.~\ref{plots}.

Although we use the same two values of $N$ for all the signals we consider for fair comparison, normally the value of $N$ should be chosen according to the extent of the signals in both the time/space and frequency domains. The error is primarily determined by how much of the signal falls outside of the extents implied by the chosen value of $N$. For example, for F1, which has a very rapidly decaying Gaussian envelope, very little falls outside so the errors are much smaller than for the others. In those cases where the results are not sufficiently accurate for the application at hand, it is possible to obtain higher accuracy by increasing N.

\begin{table*}[!t]
\renewcommand{\arraystretch}{1.4}
\centering
\caption{Percentage MSE Errors for Different Functions and Transforms (for both ordinary and centered schemes)}
\label{mse_scores_table_ord}
\begin{tabular}{llllllllll}
\hline
Input $\:$ N & T1 (ord.) & T2 (ord.) & T3 (ord.) & T4 (ord.) & T1 (cent.)& T2 (cent.)& T3 (cent.)& T4 (cent.) \\
\hline
\multirow{2}{*}{F1}  $\:$  $\:$  256  &$9.82\!\! \times \!\!10^{-4}$&$4.72 \!\!\times \!\! 10^{-3}$&$6.78 \!\! \times \!\! 10^{-4}$&$3.93 \!\! \times \!\!10^{-2}$ &$9.82 \!\!\times\!\! 10^{-4}$&$4.71 \!\!\times \!\!10^{-3}$&$6.78 \!\!\times \!\! 10^{-4}$&$3.93 \!\! \times \!\!10^{-2}$\\
                    $\:$ $\:$ $\:$ $\:$ 1024  &$6.40 \!\!\times \!\! 10^{-5}$&$2.76 \!\!\times\!\! 10^{-4}$&$4.26 \!\!\times\!\! 10^{-5}$&$2.49 \!\!\times\!\! 10^{-3}$ &$6.40 \!\!\times \!\!10^{-5}$&$2.76\!\! \times\!\! 10^{-4}$&$4.26\!\! \times\!\! 10^{-5}$&$2.49\!\! \times\!\! 10^{-3}$\\
\hline
\multirow{2}{*}{F2}  $\:$  $\:$  256  &$4.31$&$10.6$&$1.95$&$6.65$&$4.31$&$10.6$&$1.96$&$6.65$\\
                    $\:$ $\:$ $\:$ $\:$ 1024   &$0.32$&$0.87$&$0.13$&$0.46$&$0.32$&$0.87$&$0.13$&$0.46$\\
  
\hline                    
\multirow{2}{*}{F3}  $\:$  $\:$  256  &$2.49$&$1.55$&$2.84$&$2.85$&$2.02$&$1.45$&$2.37$&$2.66$\\
               $\:$ $\:$ $\:$ $\:$ 1024    &$1.09$&$0.75$&$1.40$&$1.44$&$1.10$&$0.85$&$1.34$&$1.50$\\
\hline                    
\multirow{2}{*}{F4}  $\:$  $\:$  256   &$1.34$&$0.64$&$2.29$&$6.77$&$1.35$&$0.63$&$2.30$&$6.79$\\
              $\:$ $\:$ $\:$ $\:$ 1024  &$9.43 \!\!\times\!\! 10^{-2}$&$4.38\!\! \times\!\! 10^{-2}$&$0.16$&$0.49$&$9.44\!\! \times\!\! 10^{-2}$&$4.38\! \!\times\!\! 10^{-2}$&$0.16$&$0.49$\\

\hline
\end{tabular}
\end{table*}

\subsection{Concatenation}

In order to test how well the concatenation property is satisfied, we employ the following procedure. Let us consider T1 and T2 as an example: First derive the DLCT matrices $\mathbf{C}_{\mathbf{L}_1}$ and
$\mathbf{C}_{\mathbf{L}_2}$ for T1 and T2 separately, following the
procedure given in Section~\ref{dlct}. Then, by using
Eq.~\ref{LCTmatrix}, we calculate the $2\times 2$ LCT parameter matrices
$\mathbf{L}_1$ and $\mathbf{L}_2$ for T1 and T2.  Multiplying these two
matrices by using Eq.~\ref{index_add_prop}, we obtain the $2\times 2$
parameter matrix of the concatenated system $\mathbf{L}_{12}=
\mathbf{L}_2\mathbf{L}_1$. Then, we obtain
$\mathbf{C}_{\mathbf{L}_{12}}$ from $\mathbf{L}_{12}$, again by using our 
proposed DLCT procedure. Finally, we compare the result of applying
the concatenated transform matrix $\mathbf{C}_\mathbf{L_{12}}$ directly
with the result of applying $\mathbf{C}_{\mathbf{L}_1}$ and
$\mathbf{C}_{\mathbf{L}_2}$ consecutively.
More precisely, we compare $\mathbf{C}_{\mathbf{L}_{12}}\mathbf{x}$ with
$\mathbf{C}_{\mathbf{L}_2}\mathbf{C}_{\mathbf{L}_1}\mathbf{x}$ where
a signal $x[n]$ of length $N$ is represented by the column vector
$\mathbf{x}$. The resulting MSE differences are tabulated in
Table~\ref{mse_scores_conca_inv} for several such concatenations among T1, T2, T3, and T4. The ordinary sampling scheme is used in
these numerical calculations.

\begin{table*}[!t]
\renewcommand{\arraystretch}{1.4}
\centering
\caption{Percentage MSE Errors for Different Concatenations and Inverses}
\label{mse_scores_conca_inv}
\begin{tabular}{llllllll}
\hline
           Input    & $\:$ $\:$ N  & T1-T2 & T3-T4 & T3-T1 & T3-T2 & T1-T1$^{-1}$ & T3-T3$^{-1}$  \\
\hline
\multirow{2}{*}{F1} & \vline $\:$  256  &$1.32 \times 10^{-2} $&$2.78 \times 10^{-3}  $&$1.55 \times 10^{-3}  $&$4.10 \times 10^{-3} $ &$5.85 \times 10^{-3} $ &$9.64 \times 10^{-4} $\\
                    & \vline $\:$ 1024  &$6.82 \times 10^{-4}  $&$1.71 \times 10^{-4} $&$9.58 \times 10^{-5} $&$2.79 \times 10^{-4}  $ &$3.85 \times 10^{-4} $ &$6.29 \times 10^{-5} $\\

\hline

\multirow{2}{*}{F2} & \vline $\:$ 256  &$17.7  $&$0.34 $&$0.35 $&$2.99  $ &$1.77 $ &$0.49 $\\
                    & \vline $\:$ 1024  &$1.64  $&$2.47 \times 10^{-2} $&$2.43 \times 10^{-2} $&$0.23  $ &$0.11 $ &$3.48 \times 10^{-2} $\\
  
\hline                    
\multirow{2}{*}{F3} & \vline $\:$ 256  &$1.47  $&$1.32 $&$0.99 $&$1.26  $ &$6.22 $ &$5.31 $\\
                    & \vline $\:$ 1024  &$1.14  $&$1.05 $&$1.01 $&$1.26  $ &$5.67 $ &$4.16 $\\
\hline                    
\multirow{2}{*}{F4} & \vline $\:$ 256  &$6.73  $&$1.77 $&$1.03 $&$2.15  $ &$18.37 $ &$1.83 $\\
                    & \vline $\:$ 1024  &$0.28  $&$0.14 $&$8.16 \times 10^{-2} $&$0.17  $ &$2.12 $ &$0.23 $\\

\hline
\end{tabular}
\end{table*}

\subsection{Reversibility}
To test the reversibility property numerically, we follow a similar
procedure as in concatenation. This time the second LCTs in the cascade are the inverses of the first ones. For example, we compare $\mathbf{x}$ with
$\mathbf{C}_{\mathbf{L}_1^{-1}}
\mathbf{C}_{\mathbf{L}_1}\mathbf{x}$. Again the ordinary sampling scheme 
is used in these calculations and the resulting MSE differences are
tabulated in Table~\ref{mse_scores_conca_inv}.

\section{Conclusion}
\label{conc}
In this paper, a definition of the discrete linear canonical transform
(DLCT) based on hyperdifferential operator theory is proposed. 
For finite-length signals of a discrete variable, a unitary DLCT
matrix is obtained so that the LCT-transformed version of the input
signal can be obtained by direct matrix multiplication. Given a
vector holding the samples of a continuous-time signal, this DLCT
matrix multiplies the vector to obtain the approximate samples of the
continuous-time LCT-transformed signal, similar to the DFT being used
to approximate the continuous-time Fourier transform.

The advantage of a discrete transform is that it provides a basis for 
numerical computation. However, our expectations were more than that. The main goal of this work was to obtain a formulation of the discrete LCT based on self-consistent definitions of the discrete coordinate multiplication and
differentiation operators, that mirror the structure of their
continuous counterparts. Care was taken to ensure that the discrete
coordinate multiplication and differentiation operators were strictly
duals of each other, related through the DFT. The resulting DLCT matrix
is totally compatible with the theory of the discrete Fourier
transform (DFT) and its dual and circulant structure.
Desirable properties of a discrete LCT definition such as unitarity,
preservation of group structure, reversibility and approximation of
the continuous LCT were discussed both theoretically and numerically. One immediate possibility for future work is to explore the application of the method to alternative decompositions, such as those discussed in \cite{koc08ieee,hennelly05fastLCT2,hennelly05fastLCT2}.

We showed in \cite{koc08ieee}, that we could digitally compute the continuous LCT to an accuracy limited by the uncertainty relationship, with a fast algorithm. However, this numerical computation method did not exhibit properties we desire from a discrete definition. On the other hand, without a fast algorithm, application of the definition proposed in the present paper involves a matrix multiplication and thus has complexity $O(N^2)$. The best of both worlds would be to find a fast algorithm for the definition proposed in the present paper. This would be analogous to first defining the DFT and then deriving the FFT algorithm for its fast computation. However, such an algorithm is presently not available and will require future work. In the meantime, fast computational methods as in \cite{koc08ieee,hennelly05fastLCT,hennelly05fastLCT2,healy10LCTalgoreview} can be used in practical applications when speed is important. The computational complexity of taking the DLCT of signals, which is a matrix multiplication with $O(N^2)$ complexity, should not be confused with the complexity of constructing the proposed DLCT matrix, which has to be done once for a particular LCT. The latter is discussed in Appendix~\ref{appendix_computationalcost}.

In the present paper our emphasis was to define the DLCT in a manner that preserves structural similarity with the continuous DLCT. The structure in question is how the LCT is defined in terms of coordinate multiplication and differentiation in terms of hyperdifferential operators, which we followed closely. Since everything rests on these two operators, their accuracy is what defines the accuracy of the method. We chose the conceptually simplest first-order approximations for these. Accuracy can be increased either by increasing $N$, or by replacing these building blocks with higher-order approximations. Thus, the hyperdifferential formulation provided here constitutes not only a theoretically pure approach to defining the DLCT, it serves as a framework for high
accuracy numerical computations.

In conclusion, we have applied hyperdifferential operator theory to
the task of defining the discrete LCT in a manner that is fully
consistent with the dual and circulant structure of the DFT. Although several definitions for the DLCT have been proposed, a comprehensive evaluation of their relationships remains an important subject for future work. We
believe our proposed analytical approach can lead to further
possible research directions in the theory of discrete transforms in general.


%

\appendices
\section{Proof of Unitarity}
\label{appendix_unitarity}

We start with $\mathbf{U}$ given in Eq.~\ref{Umatrix}. $\mathbf{U}$ is
a real diagonal matrix, which implies it is Hermitian. The next step
is to show $\mathbf{D}$ is also Hermitian. Starting from
Eq.~\ref{D_fuf}, we can write
\begin{eqnarray}
\mathbf{D^H} = \mathbf{(F^{-1}UF)^H} = \mathbf{F^H U^H(F^H)^H} = \mathbf{F^{-1} UF} = \mathbf{D} \nonumber 
\end{eqnarray}
implying that $\mathbf{D}$ is also Hermitian. Now, we move on to show
that $\mathbf{Q}_q$, $\mathbf{M}_M$, and $\mathbf{F}_{\text{lc}}^a$
are unitary given $\mathbf{U}$ and $\mathbf{D}$ are Hermitian, by showing that their inverses and their Hermitians are equal. The inverse of $\mathbf{Q}_q$ is
\begin{equation}
\label{Qinv}
\mathbf{Q}^{-1}_q=\mathbf{Q}_{-q}=\exp{\left(i2\pi q \,
\frac{\mathbf{U}^2}{2}\right)}
\end{equation}
while the Hermitian of $\mathbf{Q}_q$ is
\begin{eqnarray}
\label{Qhermitian}
\mathbf{Q}_q^\mathbf{H}=\exp{\left(i2\pi q \,
\frac{\mathbf{(U^H})^2}{2}\right)} 
=\exp{\left(i2\pi q \,
\frac{\mathbf{U}^2}{2}\right)},
\end{eqnarray}
which are equal to each other. Similarly, one can follow the same procedure for $\mathbf{M}_M$ as follows:

\begin{eqnarray}
\label{Minv}
\mathbf{M}^{-1}_M&=&\mathbf{M}_{1/M}=\exp{\left(-i2\pi \ln{(1/M)}\,
\frac{\mathbf{UD+DU}}{2}\right)} \nonumber \\
&=&\exp{\left(i2\pi \ln{(M)}\,
\frac{\mathbf{UD+DU}}{2}\right)}
\end{eqnarray}
and
\begin{eqnarray}
\label{Mhermitian}
\mathbf{M}_M^\mathbf{H} &=&\exp{\left(i2\pi \ln{(M)}\,
\frac{\mathbf{(UD+DU)^H}}{2}\right)} \nonumber \\
&=&\exp{\left(i2\pi \ln{(M)}\,
\frac{\mathbf{DU+UD}}{2}\right)} 
=\mathbf{M}^{-1}_M.
\end{eqnarray}
And, finally for $\mathbf{F}_{\text{lc}}^a$ we can write:

\begin{equation}
\label{Finv}
(\mathbf{F}^a_{\text{lc}})^{-1}=\mathbf{F}^{-a}_{\text{lc}}=\exp{\left(ia\pi^2 \,
\frac{\mathbf{U}^2+\mathbf{D}^2}{2}\right)} 
\end{equation}
and
\begin{eqnarray}
\label{Fhermitian}
(\mathbf{F}^a_{\text{lc}})^{\mathbf{H}}&=&\exp{\left(ia\pi^2 \,
\frac{(\mathbf{U}^2+\mathbf{D}^2)^\mathbf{H}}{2}\right)} 
=  (\mathbf{F}^a_{\text{lc}})^{-1}.
\end{eqnarray}
The first equalities in Eqs.~\ref{Qhermitian}, \ref{Mhermitian}, and \ref{Fhermitian} can be shown by considering power expansion formula (Appendix \ref{appendix_fundamentals}). Thus we have proven Theorem \ref{theorem2} and therefore Theorem \ref{theorem1}. Justifications for the intermediate steps above will be given in the Appendix~\ref{appendix_fundamentals}.

\section{Some Fundamentals of Operator Theory}
\label{appendix_fundamentals}

Here we provide further details regarding the derivations that appear in Section \ref{dlct} and Appendix \ref{appendix_unitarity}. These derivations are mostly based on the following elementary definitions or results: (i) The integer power of an operator is defined as its repeated application, e.g. ${\cal A}^3 = {\cal A}{\cal A}{\cal A}$. (ii) Therefore, any power of ${\cal A}$ commutes with itself, i.e. ${\cal A}^n{\cal A}$= ${\cal A}{\cal A}^n$. (iii) This leads to the fact that any polynomial $p({\cal A})$ of ${\cal A}$ commutes with ${\cal A}$, i.e. $p({\cal A}){\cal A} = {\cal A}p({\cal A})$. (iv) Functions such as $\exp({\cal A})$ and $\sin({\cal A})$ can be defined through power series of $\exp(\cdot)$ and $\sin(\cdot)$, which are essentially like polynomials, therefore these functions of ${\cal A}$ also commute with ${\cal A}$. (v) Carrying this one step further, two different functions of ${\cal A}$ that can be expressed as power series will also commute with each other, again as a consequence of (ii). (vi) The Hermitian of $p({\cal A})$, and thus also $\exp({\cal A})$ and $\sin({\cal A})$ can be obtained by replacing ${\cal A}$ with its Hermitian inside the power series. This follows from the fact that $({\cal A}^n)^{\rm H} = ({\cal A}^{\rm H})^n$.

Eq.~\ref{sincUU} follows directly from (iv) above. Eq.~\ref{U_final_op} follows from the fact that the effect of ${\cal U}$ on a continuous signal $f(u)$ is to multiply it with $u$, and the fact that $\sin({\cal U})$ can be written as a power series of ${\cal U}$.

The steps in Eqs.~\ref{Qinv} to \ref{Fhermitian} in the Appendix~\ref{appendix_unitarity} are most clearly established as follows. For the first equality in Eq.~\ref{Qhermitian}, it follows from (vi) in the established facts above. With regards to Eq.~\ref{Qinv}, we observe that Eqs.~\ref{chirpmatrix} and \ref{chirpcont} show that the inverse of the chirp multiplication operator is again a similar operator but with negative parameter. Similar observations can be made for the other operators by referring to their $2\times 2$ matrices.  Regarding Eq.~\ref{Qinv}, this means that the inverse of a chirp multiplication operator is of the same form but with negative parameter $-q$. So we need to show that
$\exp(i2\pi q{\bf U}^2/2) \exp(-i2\pi q{\bf U}^2/2)
$
is equal to the identity. Here we can invoke the Baker-Campbell-Hausdorff formula for matrices, \cite{hall03glauber,cohentannoudji77quantum}, which states that
\begin{equation}
\exp({\bf A})\exp({\bf B}) = \exp({\bf A}+{\bf B} + 1/2({\bf A}{\bf B}-{\bf B}{\bf A})),
\end{equation}
for two complex matrices ${\bf A}$ and ${\bf B}$ where both ${\bf A}$ and ${\bf B}$ commute with their commutator $({\bf A}{\bf B}-{\bf B}{\bf A})$.

In our case, ${\bf A} = -{\bf B}$, so that $({\bf A}{\bf B}-{\bf B}{\bf A})={\bf 0}$. Therefore, the Baker-Campbell-Hausdorff formula's condition is met since every matrix commutes with the zero matrix. Finally, we observe that the product on the left-hand side of the above identity becomes equal to the exponential of the zero matrix and therefore the identity operator, proving the claim. Exactly the same argument applies for Eq.~\ref{Minv} and Eq.~\ref{Finv} since, although the exponents are more complicated, in each case a minus sign is introduced to the exponent but otherwise the exponent remains the same. Therefore the exponent of the original and the inverse are merely negatives of each other and will commute, so that the product of the original and inverse matrices will be the identity.

The ${\rm sinc}(x)=\sin(\pi x)/(\pi x)$ function has a power series that is obtained by dividing the power series of $\sin(\pi x)$ by $(\pi x)$. From number (iv) of our elementary results, ${\rm sinc}(h{\cal D})$ commutes with ${\cal D}$, so both forms in Eq.~\ref{D_hyp} are the same. The same is true for Eq.~\ref{sincUU}.

\section{Computation of the Matrix Exponential}
\label{expm}

Although it may be viewed as an implementation detail, given that it
lies at the heart of the proposed method, it is worth clarifying how
to compute the matrix exponential operation in Eq.~\ref{lct_full}.
In practice, it is common to use \textit{MATLAB}'s standard routines
to compute matrix exponentials. Mathematically, the way in which
matrix exponentials are obtained is through the well-known eigen
decomposition
\begin{equation}
\mathbf{A} = \mathbf{PDP^{-1}}
\end{equation}
where $\mathbf{D}$ is a diagonal matrix that holds the eigenvalues of
$\mathbf{A}$ and $\mathbf{P}$ is the matrix holding the
eigenvectors. Then, $\exp(\mathbf{A})=\mathbf{P\exp(D)P^{-1}}$ where the
$\exp()$ that operates on $\mathbf{D}$ is now simply an element-wise
exponentiation operation. When $\mathbf{A}$ has a full set of eigenvalues,
this procedure works without any complication. Given
Eqs.~\ref{Umatrix} and~\ref{D_fuf}, and the unitarity of the DFT matrix $\mathbf{F}$,
the matrices $\mathbf{U}$ and $\mathbf{D}$ are ensured to have a full
set of eigenvalues and eigenvectors, so there is no mathematical
complication in using matrix exponentials.

\section{Computational Cost of Constructing the Proposed DLCT Matrix}
\label{appendix_computationalcost}
Given a specified precision (i.e., number of bits used in computations is fixed), to find the complexity of generating the matrix $\mathbf{C}_\mathbf{L}$ as a function of $N$, we first find the complexity of computing the matrices $\mathbf{U}$ and $\mathbf{D}$. The matrix $\mathbf{U}$ is generated using Eq. 33. This process requires evaluation of the sine function at $N$ points and $N$ multiplications by the constant $\sqrt{N}/\pi$. Since we assume a fixed precision, we can take the evaluation of the sine function at a point to be of complexity $O(1)$. The complexity of computing $\mathbf{U}$ is thus $O(N)$. Secondly, to compute $\mathbf{D}$ using Eq. 34, we need to compute the matrix $\mathbf{F}$ and $\mathbf{F}^{-1}$, both of which can be written in terms of $W_{N}$. In generating $\mathbf{F}$, we compute $W_{N}$ only once and compute its $(mn)$'th power for the $(mn)$'th entry. Computing the $(mn)$'th entry for the matrices $\mathbf{F}$ and $\mathbf{F}^{-1}$ requires two multiplications and one exponentiation, which are each taken to be $O(1)$. It follows that computing $\mathbf{F}$ and $\mathbf{F}^{-1}$ each takes $O(N^2)$ computations. Finally, multiplying $\mathbf{F}^{-1}$ with $\mathbf{U}$ is $O(N^2)$ since $\mathbf{U}$ is diagonal whereas multiplying $\mathbf{F}^{-1} \mathbf{U}$ with $\mathbf{F}$ is $O(N^2 \log N)$ (by using fast Fourier transform (FFT) algorithm and by noting that neither matrices are diagonal), resulting in an overall complexity of $O(N^2 \log N)$ for $\mathbf{D}$.

We can now move on to the complexities of computing the matrices $\mathbf{Q}_q, \mathbf{M}_M, \mathbf{F}_{\text{lc}}^a$ based on Eqs. 23, 24, and 25. Note that in Eqs. 23, 24, and 25, the scalar constants can be taken outside the $\exp()$ function, be computed separately and then be multiplied with the resulting matrix exponentials. This does not have an effect on the computational complexity with respect to $N$. 

\begin{itemize}
    \item Complexity of $\mathbf{Q}_q$: Taking the square of $\mathbf{U}$ is of complexity $O(N)$ since $\mathbf{U}$ is a diagonal matrix. We can compute the matrix exponential of $\mathbf{U}^2$ simply by taking the exponential of each diagonal element because $\mathbf{U}^2$ is also a diagonal matrix. This amounts to an overall computational complexity of $O(N)$.
    
    \item Complexity of $\mathbf{M}_M$: One can compute both $\mathbf{U}\mathbf{D}$ and $\mathbf{D}\mathbf{U}$ in $O(N^2)$ time because $\mathbf{U}$ is a diagonal matrix. However, generating $\mathbf{D}$ increases the time to compute the argument of the $\exp()$ to $O(N^2 \log N)$. Furthermore, computing matrix exponentials as described in Appendix C is of complexity $O(N^3)$. As a result, the overall complexity is $O(N^3)$.
    
    \item Complexity of $\mathbf{F}_{\text{lc}}^a$: This is the same as the complexity of $\mathbf{M}_M$ since it involves computing the matrix exponential of a non-diagonal matrix.
\end{itemize}

In conclusion, the overall complexity for computing the matrix $\mathbf{C}_\mathbf{L}$ is $O(N^3)$.


\bibliographystyle{plain}

\bibliography{archives}

\begin{thebibliography}{10}

\bibitem{koc17eswalct}
Sparse representation of two- and three-dimensional images with fractional
  {F}ourier, {H}artley, linear canonical, and {H}aar wavelet transforms.
\newblock {\em Expert Systems with Applications}, 77:247 -- 255, 2017.

\bibitem{abe94mainlct}
S.~Abe and J.~T. Sheridan.
\newblock Generalization of the fractional {F}ourier transformation to an
  arbitrary linear lossless transformation an operator approach.
\newblock {\em Journal of Physics A: Mathematical and General},
  27(12):4179--4187, 1994.

\bibitem{abe94abcd}
S.~Abe and J.~T. Sheridan.
\newblock Optical operations on wavefunctions as the abelian subgroups of the
  special affine {F}ourier transformation.
\newblock {\em Opt. Lett.}, 19:1801--1803, 1994.

\bibitem{alieva07lctproperties}
T.~Alieva and M.~J. Bastiaans.
\newblock Properties of the canonical integral transformation.
\newblock {\em J. Opt. Soc. Am. A}, 24:3658--3665, 2007.

\bibitem{atakishiyev99dfrt}
N.~M. Atakishiyev, L.~E. Vicent, and K.~B. Wolf.
\newblock Continuous vs. discrete fractional {F}ourier transforms.
\newblock {\em Journal of Computational and Applied Mathematics}, 107(1):73 --
  95, 1999.

\bibitem{barker00dfrt}
Laurence Barker, Cagatay Candan, Tugrul Hakioglu, M~Alper Kutay, and Haldun~M
  Ozaktas.
\newblock The discrete harmonic oscillator, {H}arper's equation, and the
  discrete fractional {F}ourier transform.
\newblock {\em Journal of Physics A: Mathematical and General}, 33(11):2209,
  2000.

\bibitem{barshan97optfilt}
B.~Barshan, M.~A. Kutay, and H.~M. Ozaktas.
\newblock Optimal filtering with linear canonical transformations.
\newblock {\em Optics Communications}, 135(1-3):32 -- 36, 1997.

\bibitem{bastiaans78wigner}
M.~J. Bastiaans.
\newblock The {W}igner distribution function applied to optical signals and
  systems.
\newblock {\em Optics Communications}, 25(1):26 -- 30, 1978.

\bibitem{bastiaans79wigner}
M.~J. Bastiaans.
\newblock Wigner distribution function and its application to first-order
  optics.
\newblock {\em J. Opt. Soc. Am.}, 69:1710--1716, 1979.

\bibitem{bastiaans07abcdclassification}
M.~J. Bastiaans and T.~Alieva.
\newblock Classification of lossless first-order optical systems and the linear
  canonical transformation.
\newblock {\em J. Opt. Soc. Am. A}, 24:1053--1062, 2007.

\bibitem{bernardo97cFRTtalbot}
L.~M. Bernardo.
\newblock Talbot self-imaging in fractional {F}ourier planes of real and
  complex orders.
\newblock {\em Optics Communications}, 140:195--198, 1997.

\bibitem{bernardo96cFRT}
L.~M. Bernardo and O.~D.~D. Soares.
\newblock Optical fractional {F}ourier transforms with complex orders.
\newblock {\em Appl. Opt.}, 35(17):3163--3166, 1996.

\bibitem{campos11fastLCT}
R.~G. Campos and J.~Figueroa.
\newblock A fast algorithm for the linear canonical transform.
\newblock {\em Signal Processing}, 91(6):1444--1447, 2011.

\bibitem{candan00dfrt}
C.~Candan, M.~A. Kutay, and H.~M. Ozaktas.
\newblock The discrete fractional {F}ourier transform.
\newblock {\em Signal Processing, IEEE Transactions on}, 48(5):1329 --1337,
  2000.

\bibitem{chen15radar}
X.~Chen, J.~Guan, Y.~Huang, N.~Liu, and Y.~He.
\newblock Radon-linear canonical ambiguity function-based detection and
  estimation method for marine target with micromotion.
\newblock {\em IEEE Trans. Geoscience and Remote Sens.}, 53(4):2225--2240,
  2015.

\bibitem{chen14radar}
X.~Chen, J.~Guan, N.~Liu, W.~Zhou, and Y.~He.
\newblock Detection of a low observable sea-surface target with micromotion via
  the radon-linear canonical transform.
\newblock {\em IEEE Geosci. Remote Sensing Lett.}, 11(7):1225--1229, 2014.

\bibitem{clary03shiftedFourier}
Stuart Clary and Dale~H. Mugler.
\newblock Shifted {F}ourier matrices and their tridiagonal commutors.
\newblock {\em SIAM Journal on Matrix Analysis and Applications},
  24(3):809--821, 2003.

\bibitem{cohentannoudji77quantum}
C.~Cohen-Tannoudji, B.~Diu, and F.~Laloe.
\newblock {\em Quantum Mechanics}.
\newblock New York: Wiley, 1979.

\bibitem{davies78integtransappl}
B.~Davies.
\newblock {\em Integral Transforms and Their Applications}.
\newblock Springer, New York, 1978.

\bibitem{erseghe99dfrt}
T.~Erseghe, P.~Kraniauskas, and G.~Carioraro.
\newblock Unified fractional {F}ourier transform and sampling theorem.
\newblock {\em Signal Processing, IEEE Transactions on}, 47(12):3419 --3423,
  dec. 1999.

\bibitem{feng162DLCT}
Qiang Feng and Bing-Zhao Li.
\newblock Convolution and correlation theorems for the two-dimensional linear
  canonical transform and its applications.
\newblock {\em IET Signal Processing}, 10:125--132, 2016.

\bibitem{grunbaum82centeredDFT}
F.Alberto Grunbaum.
\newblock The eigenvectors of the discrete {F}ourier transform: A version of
  the {H}ermite functions.
\newblock {\em Journal of Mathematical Analysis and Applications}, 88(2):355 --
  363, 1982.

\bibitem{gulcu18choiceQ}
Talha~Cihad Gulcu and Haldun~M. Ozaktas.
\newblock Choice of quantization interval for finite-energy fields.
\newblock {\em IEEE Transactions on Signal Processing}, 66:2470--2479, 2018.

\bibitem{hall03glauber}
B.~C. Hall.
\newblock {\em Lie Groups, Lie Algebras, and Representations (Chapter~3: The
  Baker—Campbell—Hausdorff Formula)}.
\newblock New York: Springer, 2003.

\bibitem{ozaktas16lctbook}
J.~J. Healy, M.~A. Kutay, H.~M. Ozaktas, and J.~T.~Sheridan eds.
\newblock {\em Linear Canonical Transforms: Theory and Applications}.
\newblock Springer New York, New York, NY, 2016.

\bibitem{healy09lctsampling}
J.~J. Healy and J.~T. Sheridan.
\newblock Sampling and discretization of the linear canonical transform.
\newblock {\em Signal Process.}, 89(4):641--648, 2009.

\bibitem{healy10LCTalgoreview}
J.~J. Healy and J.~T. Sheridan.
\newblock Fast linear canonical transforms.
\newblock {\em J. Opt. Soc. Am. A}, 27(1):21--30, 2010.

\bibitem{healy10reevaluation}
John~J. Healy and John~T. Sheridan.
\newblock Reevaluation of the direct method of calculating {F}resnel and other
  linear canonical transforms.
\newblock {\em Opt. Lett.}, 35(7):947--949, Apr 2010.

\bibitem{hecht01book}
E.~Hecht.
\newblock {\em Optics, 4th Ed.}
\newblock Addison Wesley, 2001.

\bibitem{hennelly05fastLCT}
B.~M. Hennelly and J.~T. Sheridan.
\newblock Fast numerical algorithm for the linear canonical transform.
\newblock {\em J. Opt. Soc. Am. A}, 22:928--937, 2005.

\bibitem{hennelly05fastLCT2}
B.~M. Hennelly and J.~T. Sheridan.
\newblock Generalizing, optimizing, and inventing numerical algorithms for the
  fractional {F}ourier, {F}resnel, and linear canonical transforms.
\newblock {\em J. Opt. Soc. Am. A}, 22:917--927, 2005.

\bibitem{hua97lct}
J.~Hua, L.~Liu, and G.~Li.
\newblock Extended fractional {F}ourier transforms.
\newblock {\em J. Opt. Soc. Am. A}, 14(12):3316--3322, 1997.

\bibitem{james96genFresnel}
D.~F.~V. James and G.~S. Agarwal.
\newblock The generalized {F}resnel transform and its application to optics.
\newblock {\em Optics Communications}, 126(4-6):207 -- 212, 1996.

\bibitem{jung82quantcLCT}
C.~Jung and H.~Kruger.
\newblock Representation of quantum mechanical wavefunctions by complex valued
  extensions of classical canonical transformation generators.
\newblock {\em J. Phys. A: Math. Gen.}, 15:3509--3523, 1982.

\bibitem{knapp01bookgroup}
A.~W. Knapp.
\newblock {\em Representation theory of semisimple groups: An overview based on
  examples}.
\newblock Princeton University Press, 2001.

\bibitem{koc18scaling}
A.~Ko\c{c}, B.~Bartan, and H.~M. Ozaktas.
\newblock Discrete scaling based on operator theory.
\newblock {\em arXiv preprint arXiv:1805.03500}, 2018.

\bibitem{koc16lctchapter}
A.~Ko\c{c}, F.~S. Oktem, H.~M. Ozaktas, and M.~A. Kutay.
\newblock {\em Linear Canonical Transforms: Theory and Applications}, chapter
  Fast Algorithms for Digital Computation of Linear Canonical Transforms, pages
  293--327.
\newblock Springer New York, New York, NY, 2016.

\bibitem{koc08ieee}
A.~Ko\c{c}, H.~M. Ozaktas, C.~Candan, and M.~A. Kutay.
\newblock Digital computation of linear canonical transforms.
\newblock {\em IEEE Trans. Signal Process.}, 56(6):2383--2394, 2008.

\bibitem{koc10complexLCT}
A.~Ko\c{c}, H.~M. Ozaktas, and L.~Hesselink.
\newblock Fast and accurate algorithm for the computation of complex linear
  canonical transforms.
\newblock {\em J. Opt. Soc. Am. A}, 27(9):1896--1908, 2010.

\bibitem{koc10nonsep2D}
A.~Ko\c{c}, H.~M. Ozaktas, and L.~Hesselink.
\newblock Fast and accurate computation of two-dimensional non-separable
  quadratic-phase integrals.
\newblock {\em J. Opt. Soc. Am. A}, 27(6):1288--1302, 2010.

\bibitem{li14watermark}
B.Z. Li and Y.P. Shi.
\newblock Image watermarking in the linear canonical transform domain.
\newblock {\em Mathematical Problems in Engineering}, 2014.

\bibitem{moshinsky73quantLCT}
M.~Moshinsky.
\newblock Canonical transformations and quantum mechanics.
\newblock {\em SIAM J. Appl. Math.}, 25(2):193--212, 1973.

\bibitem{moshinsky71lct}
M.~Moshinsky and C.~Quesne.
\newblock Linear canonical transformations and their unitary representations.
\newblock {\em Journal of Mathematical Physics}, 12(8):1772--1780, 1971.

\bibitem{mugler11centeredDFT}
D.~H. Mugler.
\newblock The centered discrete {F}ourier transform and a parallel
  implementation of the fft.
\newblock In {\em 2011 IEEE International Conference on Acoustics, Speech and
  Signal Processing (ICASSP)}, pages 1725--1728, May 2011.

\bibitem{nazarathy80operatoralgebra}
M.~Nazarathy and J.~Shamir.
\newblock Fourier optics described by operator algebra.
\newblock {\em J. Opt. Soc. Am.}, 70:150--159, 1980.

\bibitem{oktem09DLCT}
F.~S. Oktem and H.~M. Ozaktas.
\newblock Exact relation between continuous and discrete linear canonical
  transforms.
\newblock {\em Signal Processing Letters, IEEE}, 16(8):727 --730, 2009.

\bibitem{ozaktas06ol}
H.~M. Ozaktas, A.~Ko\c{c}, I.~Sari, and M.~A. Kutay.
\newblock Efficient computation of quadratic-phase integrals in optics.
\newblock {\em Opt. Lett.}, 31:35--37, 2006.

\bibitem{ozaktas01book}
H.~M. Ozaktas, Z.~Zalevsky, and M.~A. Kutay.
\newblock {\em The Fractional {F}ourier Transform with Applications in Optics
  and Signal Processing}.
\newblock New York: Wiley, 2001.

\bibitem{ozaktas10PoularikasBook}
Haldun~M. Ozaktas, M.~Alper Kutay, and Cagatay Candan.
\newblock {\em Transforms and Applications Handbook}, chapter Fractional
  {F}ourier Transform, pages 14--1--14--28.
\newblock CRC Press, Boca Raton, New York, NY, 2010.

\bibitem{ozcelikkale12finitebits}
A.~Ozcelikkale and H.~M. Ozaktas.
\newblock Representation of optical fields using finite numbers of bits.
\newblock {\em Optics letters}, 37:2193--2195, 2012.

\bibitem{ozcelikkale13randomfields}
Ayça Ozcelikkale and Haldun~M. Ozaktas.
\newblock Optimal representation of non-stationary random fields with finite
  numbers of samples: A linear {MMSE} framework.
\newblock {\em Digital Signal Processing}, 23(5):1602 -- 1609, 2013.

\bibitem{palma97mainABCD}
C.~Palma and V.~Bagini.
\newblock Extension of the {F}resnel transform to abcd systems.
\newblock {\em J. Opt. Soc. Am. A}, 14(8):1774--1779, 1997.

\bibitem{pei00DLCTdef}
S.~C. Pei and J.~J. Ding.
\newblock Closed-form discrete fractional and affine {F}ourier transforms.
\newblock {\em IEEE Trans. Signal Process.}, 48:1338--1353, 2000.

\bibitem{pei02eigenLCT}
S.~C. Pei and J.~J. Ding.
\newblock Eigenfunction of linear canonical transform.
\newblock {\em IEEE Trans. Signal Process.}, 50:11--26, 2002.

\bibitem{pei16fast_dlct}
S.~C. Pei and S.~Huang.
\newblock Fast discrete linear canonical transform based on cm-cc-cm
  decomposition and fft.
\newblock {\em IEEE Trans. Signal Process.}, 64:855 -- 866, 2016.

\bibitem{pei11dlct}
S.~C. Pei and Y.~C. Lai.
\newblock Discrete linear canonical transforms based on dilated {H}ermite
  functions.
\newblock {\em J. Opt. Soc. Am. A}, 28(8):1695 --1708, 2011.

\bibitem{pei97discretefrt}
S.~C. Pei and M.~H. Yeh.
\newblock Improved discrete fractional {F}ourier transform.
\newblock {\em Opt. Lett.}, 22(14):1047--1049, 1997.

\bibitem{pei99dfrt2}
S.~C. Pei, M.~H. Yeh, and T.~L. Luo.
\newblock Fractional {F}ourier series expansion for finite signals and dual
  extension to discrete-time fractional {F}ourier transform.
\newblock {\em Signal Processing, IEEE Transactions on}, 47(10):2883 --2888,
  1999.

\bibitem{pei99dfrt}
S.~C. Pei, M.~H. Yeh, and C.~C. Tseng.
\newblock Discrete fractional {F}ourier transform based on orthogonal
  projections.
\newblock {\em Signal Processing, IEEE Transactions on}, 47(5):1335 --1348,
  1999.

\bibitem{qi15watermark}
Min Qi, Bing-Zhao Li, and Huafei Sun.
\newblock Image watermarking using polar harmonic transform with parameters in
  {SL}(2,{R}).
\newblock {\em Signal Processing: Image Communication}, 31:161 -- 173, 2015.

\bibitem{qiu13speech}
W.~Qiu, B.Z. Li, and X.W. Li.
\newblock Speech recovery based on the linear canonical transform.
\newblock {\em Speech Communication}, 55(1):40--50, 2013.

\bibitem{rodrigo062DNSLCT}
J.~Rodrigo, T.~Alieva, and M.~Luisa Calvo.
\newblock Optical system design for orthosymplectic transformations in phase
  space.
\newblock {\em J. Opt. Soc. Am. A}, 23:2494--2500, 2006.

\bibitem{sharma09fracLaplace}
K.~K. Sharma.
\newblock Fractional {L}aplace transform.
\newblock {\em Signal, Image And Video Process.}, 4(3):377--379, 2009.

\bibitem{shih95cFRT}
C.~C. Shih.
\newblock Optical interpretation of a complex-order {F}ourier transform.
\newblock {\em Opt. Lett.}, 20(10):1178--1180, 1995.

\bibitem{siegman86book}
A.~E. Siegman.
\newblock {\em Lasers}.
\newblock Mill Valley, California: University Science Books, 1986.

\bibitem{simon00setparaxial}
R.~Simon and K.~B. Wolf.
\newblock Structure of the set of paraxial optical systems.
\newblock {\em J. Opt. Soc. Am. A}, 17(2):342--355, 2000.

\bibitem{singh10encryp}
N.~Singh and A.~Sinha.
\newblock Chaos based multiple image encryption using multiple canonical
  transforms.
\newblock {\em Optics and Laser Technology}, 42:724--731, 2010.

\bibitem{stern06whyis}
A.~Stern.
\newblock Why is the linear canonical transform so little known?
\newblock {\em IP Conf. Proc.}, pages 225--234, 2006.

\bibitem{torre02fracLaplace}
A.~Torre.
\newblock Linear and radial canonical transforms of fractional order.
\newblock {\em J. Compt. and Appl. Math.}, 153:477--486, 2003.

\bibitem{vargasrubio05centered}
J.~G. Vargas-Rubio and B.~Santhanam.
\newblock On the multiangle centered discrete fractional {F}ourier transform.
\newblock {\em IEEE Signal Process. Lett.}, 12(4):273--276, 2005.

\bibitem{wang02cFRT}
C.~Wang and B.~Lu.
\newblock Implementation of complex-order {F}ourier transforms in complex abcd
  optical systems.
\newblock {\em Optics Communications}, 203(1-2):61 -- 66, 2002.

\bibitem{wei16randomDLCT}
Deyun Wei, Ruikui Wang, and Yuan-Min Li.
\newblock Random discrete linear canonical transform.
\newblock {\em J. Opt. Soc. Am. A}, 33(12):2470--2476, Dec 2016.

\bibitem{wolf74clct1}
K.~B. Wolf.
\newblock Canonical transformations i. complex linear transforms.
\newblock {\em J. Math. Phys.}, 15(8):1295--1301, 1974.

\bibitem{wolf77clct}
K.~B. Wolf.
\newblock On self-reciprocal functions under a class of integral transforms.
\newblock {\em J. Math. Phys.}, 18(5):1046--1051, 1977.

\bibitem{wolf79book}
K.~B. Wolf.
\newblock {\em Integral Transforms in Science and Engineering (Chapter~9:
  Construction and properties of canonical transforms)}.
\newblock New York: Plenum Press, 1979.

\bibitem{wolf05frt}
K.~B. Wolf.
\newblock Finite systems, fractional {F}ourier transforms and their finite
  phase spaces.
\newblock {\em Czechoslovak Journal of Physics}, 55:1527--1534, 2005.

\bibitem{wolf16lctchapter}
K.~B. Wolf.
\newblock {\em Linear Canonical Transforms: Theory and Applications}, chapter
  Development of Linear Canonical Transforms: A Historical Sketch, pages 3--28.
\newblock Springer New York, New York, NY, 2016.

\bibitem{wolf07dfrt}
K.~B. Wolf and G.~Kr\"{o}tzsch.
\newblock Geometry and dynamics in the fractional discrete {F}ourier transform.
\newblock {\em J. Opt. Soc. Am. A}, 24(3):651--658, 2007.

\bibitem{yeh05dfrt}
M.~H. Yeh.
\newblock Angular decompositions for the discrete fractional signal transforms.
\newblock {\em Signal Processing}, 85(3):537 -- 547, 2005.

\bibitem{yetik00frtdec}
I.~S. Yetik, M.~A. Kutay, H.~Ozaktas, and H.~M. Ozaktas.
\newblock Continuous and discrete fractional {F}ourier domain decomposition.
\newblock volume~1, pages 93 --96 vol.1, 2000.

\bibitem{yosida84operationalcalc}
K.~Yosida.
\newblock {\em Operational Calculus: A Theory of Hyperfunctions}.
\newblock Springer, New York, USA, 1984.

\bibitem{zayed99frtsampling}
A.~I. Zayed and A.~G. Garc�a.
\newblock New sampling formulae for the fractional {F}ourier transform.
\newblock {\em Signal Processing}, 77(1):111 -- 114, 1999.

\bibitem{zhang13DLCT}
Feng Zhang, Ran Tao, and Yue Wang.
\newblock Discrete linear canonical transform computation by adaptive method.
\newblock {\em Opt. Express}, 21(15):18138--18151, Jul 2013.

\bibitem{zhao08rateconversionLCT}
Juan Zhao, Ran Tao, and Yue Wang.
\newblock Sampling rate conversion for linear canonical transform.
\newblock {\em Signal Processing}, 88(11):2825 -- 2832, 2008.

\bibitem{zhao13dlct}
Liang Zhao, John~J. Healy, and John~T. Sheridan.
\newblock Unitary discrete linear canonical transform: analysis and
  application.
\newblock {\em Appl. Opt.}, 52(7):C30--C36, Mar 2013.

\end{thebibliography}

\end{document}